\documentclass[twocolumn]{aastex63}

\usepackage{multirow}

\received{September 10, 2019}
\accepted{February 20, 2020}

\shorttitle{ALMA CO Observations of Long-duration GRB Hosts. I}
\shortauthors{Hatsukade et al.}

\begin{document}

\title{ALMA CO Observations of the Host Galaxies of Long-duration Gamma-ray Bursts. I: Molecular Gas Scaling Relations}

\author{Bunyo Hatsukade}
\affiliation{Institute of Astronomy, Graduate School of Science, University of Tokyo, 2-21-1 Osawa, Mitaka, Tokyo 181-0015, Japan}
\email{hatsukade@ioa.s.u-tokyo.ac.jp}

\author{Kouji Ohta}
\affiliation{Department of Astronomy, Kyoto University, Kyoto 606-8502, Japan}

\author{Tetsuya Hashimoto}
\affiliation{National Tsing Hua University, No. 101, Section 2, Kuang-Fu Road, Hsinchu, 30013, Taiwan}

\author{Kotaro Kohno}
\affiliation{Institute of Astronomy, Graduate School of Science, University of Tokyo, 2-21-1 Osawa, Mitaka, Tokyo 181-0015, Japan}
\affiliation{Research Center for the Early Universe, The University of Tokyo, 7-3-1 Hongo, Bunkyo, Tokyo 113-0033, Japan}

\author{Kouichiro Nakanishi}
\affiliation{National Astronomical Observatory of Japan, 2-21-1 Osawa, Mitaka, Tokyo 181-8588, Japan}
\affiliation{SOKENDAI (The Graduate University for Advanced Studies), 2-21-1 Osawa, Mitaka, Tokyo 181-8588, Japan}

\author{Yuu Niino}
\affiliation{Research Center for the Early Universe, The University of Tokyo, 7-3-1 Hongo, Bunkyo, Tokyo 113-0033, Japan}
\affiliation{Institute of Astronomy, Graduate School of Science, University of Tokyo, 2-21-1 Osawa, Mitaka, Tokyo 181-0015, Japan}

\author{Yoichi Tamura}
\affiliation{Division of Particle and Astrophysical Science, Graduate School of Science, Nagoya University, Nagoya 464-8602, Japan}

\begin{abstract}
We present the results of CO observations toward 14 host galaxies of long-duration gamma-ray bursts (GRBs) at $z = 0.1$--2.5 by using the Atacama Large Millimeter/submillimeter Array. 
We successfully detected CO(3--2) or CO(4--3) emission in eight hosts ($z = 0.3$--2), which more than doubles the sample size of GRB hosts with CO detection. 
The derived molecular gas mass is $M_{\rm gas} =$ (0.2--$6) \times 10^{10}$$M_{\odot}$ assuming metallicity-dependent CO-to-H$_2$ conversion factors. 
By using the largest sample of GRB hosts with molecular gas estimates (25 in total, of which 14 are CO-detected) including results from the literature, we compared molecular gas properties with those of other star-forming galaxies (SFGs). 
The GRB hosts tend to have a higher molecular gas mass fraction ($\mu_{\rm gas}$) and a shorter gas depletion timescale ($t_{\rm depl}$) as compared with other SFGs at similar redshifts especially at $z \lesssim 1$. 
This could be a common property of GRB hosts or an effect introduced by the selection of targets which are typically above the main-sequence line. 
To eliminate the effect of selection bias, we analyzed $\mu_{\rm gas}$ and $t_{\rm depl}$ as a function of the distance from the main-sequence line ($\delta$MS). 
We find that the GRB hosts follow the same scaling relations as other SFGs, where $\mu_{\rm gas}$ increases and $t_{\rm depl}$ decreases with increasing $\delta {\rm MS}$. 
No molecular gas deficit is observed when compared to other SFGs of similar SFR and stellar mass. 
These findings suggest that the same star-formation mechanism is expected to be happening in GRB hosts as in other SFGs. 
\end{abstract}


\section{Introduction} \label{sec:introduction}

Long-duration gamma-ray bursts (GRBs) have been shown to be associated with the explosions of massive stars \citep[e.g.,][]{hjor03, stan03}. 
GRBs are expected to be a new tool for probing the star-forming activity in the distant universe \citep[e.g.,][]{tota97, wije98, kist09} because i) they are related to star formation, ii) they are bright enough to be observable in the cosmological distances \citep[e.g.,][]{tanv09, salv09}, and iii) observations of afterglows and host galaxies provide information about the interstellar medium or star-forming activity in GRB environments. 
It is still a subject of debate whether GRBs can be used as unbiased tracers of star formation, that is, whether GRBs occur in normal star-forming environments. 
For example, observations of GRB hosts suggest that GRBs occur more often in low-metallicity environments \citep[e.g.,][]{stan06, leve10b, grah13, perl16}. 
Theoretical models also support this preference for low-metallicity environments, where a line-driven mass loss in the progenitor is avoided to form an accretion disk upon collapse \citep[e.g.,][]{woos06, lang06, yoon06}. 
Although multi-wavelength observations of GRB hosts have been conducted, the understanding of the properties of molecular gas, which is the fuel for star formation, has not been well provided. 
Observations of molecular gas in GRB hosts are important for establishing the link between GRBs and star-forming activity. 
Although molecular hydrogen in absorption in the spectra of GRB afterglows provides useful information in the vicinity or line of sight to GRBs \citep[e.g.,][]{proc09, kruh13, deli14}, molecular lines need to be detected in emission to measure the gas content in host galaxies.

Searches for CO line emission have been conducted to probe molecular gas in GRB hosts \citep{kohn05, endo07, hats07, hats11, stan11}. 
So far, six GRB hosts have been detected in CO emission: 
GRB~980425 at $z = 0.0085$ \citep{mich18}, GRB~051022 at $z = 0.809$ \citep{hats14}, GRB~080207 at $z = 2.0858$ \citep{arab18, mich18, hats19}, GRB~080517 at $z = 0.089$ \citep{stan15}, GRB~111005A at $z = 0.01326$ \citep{mich18}, and 190114C at $z = 0.425$ \citep{deug19}. 
Earlier works have suggested a deficiency of molecular gas in GRB hosts for their star formation rate (SFR) or stellar mass \citep{hats14, stan15, mich16}. 
\cite{stan15} found a shorter gas depletion timescale for GRB hosts compared to local star-forming galaxies, suggesting that GRBs occur toward the end of a star formation episode or in the burst phase of star formation. 
A possible scenario is that a recent merger or gas inflow induced star formation and that the progenitor formed in the star formation episode that took place in newly accreted gas, as suggested in {\sc Hi} observations \citep[e.g.,][]{mich15, arab15}. 
\cite{mich15} proposed that star formation proceeds directly in the atomic gas before it converts to the molecular phase, resulting a lower molecular gas mass for its SFR. 
However, recent studies have shown that the difference is not significant and that GRB hosts
have more diverse molecular gas properties with an additional sample, appropriate CO-to-H$_2$ conversion factors ($\alpha_{\rm CO}$), or choice of comparison sample \citep{arab18, mich18}. 
\cite{arab18} detected the CO(3--2) line in a $z = 2.0858$ host of GRB~080207 with the Plateau de Bure/NOrthern Extended Millimeter Array (NOEMA).
They found that the host was molecular gas-rich and that the molecular gas mass fraction and gas depletion timescale were comparable to those of typical star-forming galaxies at similar redshifts. 
\cite{hats19} conducted a detailed study of the molecular gas properties in the same GRB host by using the CO(1--0) line data with the Karl G. Jansky Very Large Array (VLA) and spatially-resolved CO(4--3) line data with the Atacama Large Millimeter/submillimeter Array (ALMA). 
They found that the host had molecular gas properties (such as gas fraction, gas depletion timescale, gas-to-dust ratio, location in the gas mass--SFR relation, and kinematics) similar to those of main-sequence (MS) galaxies at similar redshifts. 
\cite{mich18} performed CO(2--1) observations of seven GRB hosts with the APEX and IRAM 30-m telescopes and detected three GRB hosts (GRBs 980425, 080207, and 111005A). 
They combined the hosts with previous CO observations and found that the GRB hosts have molecular properties that are consistent with those of other galaxies. 
Recently, \cite{deug19} detected the CO(3--2) line in an interacting galaxy hosting GRB~190114C at $z = 0.425$ and found a high molecular gas fraction.

Because the molecular gas mass fraction or depletion timescale are correlated with other physical quantities such as SFR, specific SFR (sSFR $= {\rm SFR}/M_*$), stellar mass, and the distance from the MS line \citep[e.g.,][]{sain12, magd12, tacc13, genz15, tacc18}, it is important to examine the molecular gas properties of GRB hosts with these parameters. 
So far, only a handful of GRB hosts have been detected in CO, and the sample size is too small to discuss the common characteristics of GRB hosts. 
In order to understand the molecular gas properties in GRB hosts, a larger sample with CO observations is necessary.

In this paper, we present the results of CO observations toward 14 GRB hosts conducted with ALMA. 
We more than doubled the sample size of GRB hosts with CO observations, providing the largest sample for statistical studies. 
The arrangement of this paper is as follows. 
Section~\ref{sec:data} outlines the GRB hosts used in this study, the observations and data reduction, and the results. 
Section~\ref{sec:sed} describes the analysis of spectral energy distributions (SEDs) for obtaining SFRs and stellar masses of the targets.
Section~\ref{sec:properties} presents the derived physical quantities. 
In Section~\ref{sec:discussion}, we describe the molecular gas properties of the hosts in comparison with other star-forming galaxy populations and discuss the scaling relations of molecular gas for GRB hosts and star-forming galaxies. 
Our conclusions are presented in Section~\ref{sec:conclusions}. 
The spatially resolved properties of molecular gas and detailed analysis of the kinematics are presented in a separate paper. 
Throughout the paper, we adopt a cosmology with $H_0=70$ km s$^{-1}$ Mpc$^{-1}$, $\Omega_{\rm{M}}=0.3$, and $\Omega_{\Lambda}=0.7$. 
SFRs in this paper are converted to a \cite{chab03} initial mass function (IMF) from a \cite{salp55} IMF by multiplying by 0.61. 
The forms of a \cite{chab03} IMF and a \cite{krou01} IMF are similar and we do not distinguish between them.

\section{Data} \label{sec:data}

\subsection{GRB Hosts} \label{sec:targets}

\begin{figure}[t]
\begin{center}
\includegraphics[width=\linewidth]{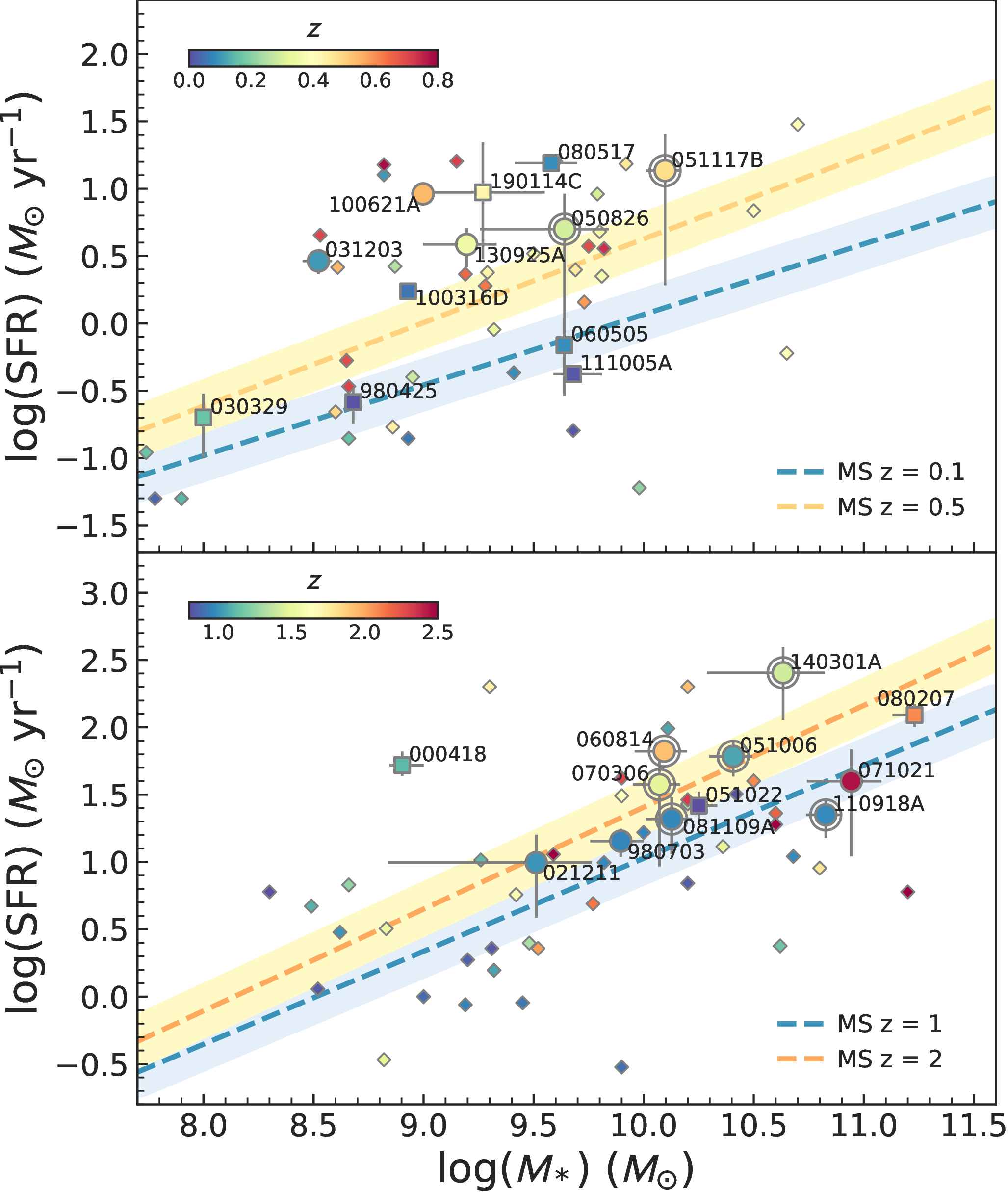}
\end{center}
\caption{
Stellar mass--SFR plot for the targets (circles), GRB hosts with CO observations in the literature (squares), and GRB hosts taken from the database of the GHostS project (diamonds). 
Top and bottom panels show the sample at redshift below and above 0.8, respectively. 
The data points and curves are color coded by redshift. 
The GRB hosts with CO detection in this study are shown as double circles. 
Curves and shaded regions represent the main sequence of star-forming galaxies at $z = 0.1, 0.5, 1.0$, and 2.0 and its uncertainty ($\pm 0.2$ dex) \citep{spea14}. 
}
\label{fig:mstar-sfr}
\end{figure}

The targets were selected from previous studies on GRB hosts, where multi-wavelength analysis are available \citep{sava09, sven10, kruh15, perl13, perl15}, thus allowing us to compare their physical quantities (such as SFR, stellar mass, or metallicity) with those of other galaxy populations. 
The selection criteria were as follows: 
(i) spectroscopic redshifts are determined to securely observe the CO lines, 
(ii) the redshifted CO line is observable with the ALMA bands, 
(iii) hosts with high SFRs ($\gtrsim$1~$M_{\odot}$~yr$^{-1}$ at $z \sim 0.1$, $\gtrsim$50~$M_{\odot}$~yr$^{-1}$ at $z \sim 1$, and $\gtrsim$150~$M_{\odot}$~yr$^{-1}$ at $z \sim 2$), 
and (iv) located at moderate redshifts ($z \le 2.5$) to ensure a significant constraint on the molecular gas mass. 
It is known that far-infrared luminosity ($L_{\rm FIR}$) correlates well with CO luminosity ($L'_{\rm CO}$) for local and high-redshift star-forming galaxies \citep[e.g.,][]{solo05}, and galaxies with high SFRs are expected to have a large amount of molecular gas. 
We selected 14 GRB hosts at $z = 0.1$--2.5 and their physical properties are presented in Table~\ref{tab:targets}. 
The SFRs adopted here are corrected for dust extinction.
It is possible that the GRB hosts with a higher radio-based SFR than a UV-based SFR have dust-obscured star-forming activity. 
The targets included dark GRBs (060814, 070306, 071021, 100621A, and GRB~130925A) whose afterglows are optically dark compared with what is expected from X-ray afterglows \citep{jako04, vand09}.

The selection criteria could introduce biases to the sample. 
GRB hosts with a spectroscopic redshift tend to have a bright afterglow or are bright at optical/near infrared (NIR) wavelengths because the redshift determination is mainly conducted through optical/NIR spectroscopic observations, which might miss heavily dust-obscured hosts. 
The SFR selection criteria target more active star-forming hosts than typical GRB hosts. 
We compare the stellar mass and SFR for the targets along with MS star-forming galaxies at $z = 0.1$--2.5, taken from \cite{spea14}, in Figure~\ref{fig:mstar-sfr}. 
A majority of the targets are located above the MS line at their redshift. 
For comparison with other GRB hosts, we used the database of the GRB Host Studies (GHostS)\footnote{\url{http://www.grbhosts.org/}} \citep{sava06}, which is a public database dedicated to GRB hosts; it contains information on more than 230 hosts. 
Figure~\ref{fig:mstar-sfr} shows that our targets have a higher SFR compared to the GHostS sample. 
In order to take the possible bias into account, we compared their properties with those of other galaxy populations by introducing the offset from the MS of star-forming galaxies (Section~\ref{sec:scaling}).

\begin{longrotatetable}
\begin{deluxetable}{ccccccccccccc}
\tablecaption{Properties of Targets} \label{tab:targets}
\tabletypesize{\small}
\tablehead{
\colhead{GRB} & \colhead{$z$} & \colhead{Ref.} & 
\colhead{SFR$_{\rm UV}$$^a$} & \colhead{Ref.} & 
\colhead{SFR$_{\rm H\alpha}$$^b$} & \colhead{Ref.} & 
\colhead{SFR$_{\rm Radio}$$^c$} & \colhead{Ref.} & 
\colhead{$M_*$$^d$} & \colhead{Ref.} & 
\colhead{$12+\log({\rm O/H})$$^e$} & \colhead{Ref.} \\
\colhead{} & \colhead{} & \colhead{} &
\colhead{($M_{\odot}$~yr$^{-1}$)} & \colhead{} &
\colhead{($M_{\odot}$~yr$^{-1}$)} & \colhead{} &
\colhead{($M_{\odot}$~yr$^{-1}$)} & \colhead{} &
\colhead{($10^9 M_{\odot}$)} & \colhead{} & 
\colhead{} & \colhead{} 
}
\startdata
980703 &$0.966\pm0.0002$ & 1&$37^{+13.1}_{-3.3}$    &13&--                  &--&$110\pm15,\ 77\pm22,\ 93\pm21$ &20,21&$5.8^{+0.4}_{-2.0}$   &13&8.15/8.31 &18 \\
021211 &$1.006\pm0.002$  & 2&$8.3^{+4.6}_{-0.7}$    &13&--                  &--&$503\pm47$, $<$41, $<$$120$    &22,23,21&$2.0^{+1.1}_{-1.0}$&13&8.29$^{f}$&-- \\
031203 &$0.1055\pm0.0001$& 3&$14.1^{+0.5}_{-0.3}$   &13&2.9                 &18&$2.9^{+0.9}_{-0.5},\ 2.3\pm0.4$&24,22&$0.3\pm0.0$           &13&8.19      &27 \\
050826 &$0.296\pm0.001$  & 4&0.85                   &14&1.8                 &18&--                             &--   &$13^{+8.3}_{-5.7}$    &18&8.48      &18 \\
051006 &$1.059\pm0.001$  & 5&$98^{+2}_{-1}$         &15&--                  &--&$51^{+22}_{-18}$               &15   &$13\pm1$              &15&8.54$^{f}$&-- \\
051117B&$0.481\pm0.001$  & 5&3.4                    &16&$4.7^{+4.9}_{-2.2}$ &19&$<$27                          &22   &20.6                  &16&8.72      &19 \\
060814 &$1.923\pm0.001$  & 6&$209^{+27}_{-53}$      &15&$54^{+89}_{-19}$    &19&$256^{+160}_{-70}$             &15   &$16^{+14}_{-6}$       &15&8.15      &21 \\
070306 &$1.496\pm0.00006$& 7&$17^{+7}_{-5}$         &15&$101^{+24}_{-18}$   &19&$143^{+61}_{-35}$              &15   &$50^{+1}_{-2}$        &15&8.36      &19 \\
071021 &$2.452\pm0.0004$ & 6&$190.3^{+25.6}_{-20.3}$&13&$32^{+20}_{-12}$    &19&--                             &--   &$119.6^{+6.6}_{-8.8}$ &13&8.12$^{g}$&-- \\
081109A&$0.9787\pm0.0005$& 8&$49.0^{+10.7}_{-10.6}$ &13&$11.8^{+4.1}_{-2.9}$&19&--                             &--   &$9.4^{+1.5}_{-1.0}$   &13&8.51      &19 \\
100621A&$0.542$          & 9&$13.5^{+5.6}_{-5.0}$   & 8&$8.7\pm0.8$         &19&$62\pm16$, $<$30               &25,26&$0.95^{+0.36}_{-0.20}$& 8&8.35      &19 \\
110918A&$0.984\pm0.001$  &10&$66^{+60}_{-30}$       &10&$41^{+28}_{-16}$    &10&$<$84                          &26   &$48^{+21}_{-15}$      &10&8.66      &19 \\
130925A&$0.347$          &11&$2.4^{+1.6}_{-1.3}$    &17&$3.1\pm1.5$         &17&--                             &--   &$3.2^{+1.9}_{-1.2}$   &17&8.50      &19 \\
140301A&$1.416$          &12&--                     &--&$106^{+36}_{-25}$   &19&--                             &--   &--                    &--&8.62      &19 \\
\enddata
\tablecomments{
$^a$ UV-based SFR from SED fitting (corrected for extinction). 
$^b$ SFR from H$\alpha$ luminosity (corrected for extinction). 
$^c$ SFR from radio continuum flux. 
$^d$ Stellar mass derived from SED fitting to UV--IR data. 
$^e$ Metallicity converted to the calibration of \cite{pett04} by using the metallicity conversion of \cite{kewl08}. 
$^f$ Metallicity derived from the mass--metallicity conversion of \cite{genz15}. 
$^g$ Metallicity derived from the line fluxes reported in \cite{kruh15}. \\
(1) \citealt{djor98}; (2) \citealt{vree03}; (3) \citealt{proc04}; (4) \citealt{mira07}; (5) \citealt{jako12}; (6) \citealt{kruh12};
(7) \citealt{jaun08}; (8) \citealt{kruh11}; (9) \citealt{milv10}; (10) \citealt{elli13}; (11) \citealt{vree13}; (12) \citealt{kruh14}; 
(13) \citealt{perl13}; (14) \citealt{sven10}; (15) \citealt{perl15}; (16) \citealt{kruh17}; (17) \citealt{scha15}; 
(18) \citealt{leve10b}; (19) \citealt{kruh15}; 
(20) \citealt{berg03}; (21) \citealt{perl17a}; (22) \citealt{mich12}; (23) \citealt{hats12}; (24) \citealt{stan10}; 
(25) \citealt{stan14}; (26) \citealt{grei16}; (27) \citealt{niin17}. 
}
\end{deluxetable}
\end{longrotatetable}

\begin{table*}[t]
\centering
\caption{ALMA Observations} \label{tab:observations}
\begin{tabular}{ccccccccccc}
\hline\hline
GRB & CO$^a$ & Band & $\nu_{\rm obs}$$^b$ & Date & Config.$^c$ & $T_{\rm on}$$^d$ & $N_{\rm ant}$$^e$ & Baseline$^f$ & $\theta$ (original)$^g$ & $\theta$ (tapered)$^h$ \\
    &     &      & (GHz)           &      &         & (min)        &               & (m)      & ($''$)          & ($''$)   \\
\hline
980703  & 4--3 & 6 & 234.507 & 2016-05-22 & C36-3   & 30 & 37 & 16.5--640.0 & $0.74 \times 0.63$ & $1.07 \times 0.99$\\
021211  & 4--3 & 6 & 229.831 & 2016-04-01 & C36-2/3 & 20 & 44 & 15.1--452.8 & $0.93 \times 0.85$ & -- \\
031203  & 3--2 & 7 & 312.796 & 2016-07-14 & C40-5   & 11 & 36 & 15.1--867.2 & $0.53 \times 0.32$ & $0.91 \times 0.75$\\
050826  & 3--2 & 6 & 266.818 & 2016-10-19 & C40-6   & 81 & 44 & 16.7--1800  & $0.19 \times 0.18$ & $0.88 \times 0.74$\\
051006  & 4--3 & 6 & 223.915 & 2016-03-31 & C36-2/3 & 24 & 44 & 15.1--452.8 & $0.99 \times 0.92$ & -- \\
051117B & 4--3 & 6 & 233.567 & 2016-10-19 & C40-6   & 90 & 44 & 16.7--1800  & $0.25 \times 0.20$ & $0.87 \times 0.75$ \\
060814  & 3--2 & 4 & 157.729 & 2016-05-04 & C36-3   & 19 & 41 & 15.1--640.0 & $1.29 \times 0.86$ & -- \\
070306  & 3--2 & 4 & 138.546 & 2016-04-28 & C36-2/3 & 44 & 41 & 15.1--640.0 & $1.24 \times 1.11$ & -- \\
071021  & 4--3 & 4 & 133.558 & 2016-05-16 & C36-3   & 31 & 41 & 15.1--640.0 & $1.52 \times 1.13$ & -- \\
081109A & 4--3 & 6 & 232.967 & 2016-05-16 & C36-3   & 19 & 40 & 15.1--640.0 & $0.68 \times 0.64$ & $1.03 \times 1.01$\\
100621A & 3--2 & 6 & 224.252 & 2016-04-01 & C36-2/3 & 21 & 44 & 15.1--452.8 & $0.98 \times 0.88$ & -- \\
\multirow{2}{*}{110918A} & \multirow{2}{*}{4--3} & \multirow{2}{*}{6} & \multirow{2}{*}{232.344} & 2016-11-05 & C40-6   & 29 & 44 & 18.6--1100  & \multirow{2}{*}{$0.44 \times 0.41$} & \multirow{2}{*}{$0.92 \times 0.86$} \\
        &      &   &         & 2016-11-27 & C40-4   & 29 & 42 & 15.1--704.1 &                &    \\
130925A & 3--2 & 6 & 256.468 & 2016-10-21 & C40-6   & 57 & 44 & 18.6--1800  & $0.22 \times 0.17$ & $0.85 \times 0.73$ \\
\multirow{2}{*}{140301A} & \multirow{2}{*}{3--2} & \multirow{2}{*}{4} & \multirow{2}{*}{143.157} & 2016-10-21 & C40-6   & 41 & 44 & 18.6--1800  & \multirow{2}{*}{$0.41 \times 0.33$} & \multirow{2}{*}{$0.85 \times 0.78$} \\
        &      &   &         & 2016-10-22 & C40-6   & 41 & 39 & 18.6--1700  &                &    \\
\hline
\multicolumn{11}{c}{\parbox{170mm}{
\vspace{1mm}
NOTE. -
$^a$ CO rotational transition. 
$^b$ Representative frequency. 
$^c$ Array configuration. 
$^d$ On-source integration time. 
$^e$ Number of antennas. 
$^f$ Range of baseline lengths. 
$^g$ Synthesized beam size in velocity-integrated CO intensity maps created with natural weighting without tapering the $uv$ data. 
$^h$ Synthesized beam size in velocity-integrated CO intensity maps created with natural weighting and $0\farcs8$ taper. 
}}
\end{tabular}
\end{table*}

In this study, we also utilized the data from CO observations in the literature: 
Six hosts with CO detection (GRBs 980425, 051022, 080207, 080517, 111005A, and 190114C) 
and five hosts with upper limits (GRBs 000418, 030329, 060505, 060814, and 100316D). 
We did not include the GRB~090423 host at $z = 8.23$ \citep{stan11} because the physical properties (such as stellar mass or SFR) are not well constrained. 
The physical properties are summarized in Table~\ref{tab:other_hosts} in Appendix~A. 
The hosts are also plotted in Figure~\ref{fig:mstar-sfr} for comparison with our targets.

\subsection{Observations and Results} \label{subsec:observations}

ALMA observations of the targets were conducted in March--November, 2016, for the Cycle 3 and Cycle 4 programs (Project code: 2015.1.00939.S and 2016.1.00455.S) as summarized in Table~\ref{tab:observations}. 
In order to observe the CO line and dust continuum simultaneously, the CO(3--2) or CO(4--3) line was observed at band 4, 6, or 7 depending on the target redshift.

The required sensitivity was estimated by using 
the SFR--$L_{\rm FIR}$ relation of \cite{kenn98}, 
and the $L_{\rm FIR}$--$L'_{\rm CO}$(3--2) relation of $L'_{\rm CO}$(3--2) $= 0.93 \times \log{L_{\rm FIR}} -1.50$ \citep{iono09}. 
The typical 5$\sigma$ detection limit for molecular gas is shown in Figure~\ref{fig:mgas-sfr} by assuming a Galactic CO-to-H$_2$ conversion factor of 4.4 $M_{\odot}$~(K~km~s$^{-1}$~pc$^2$)$^{-1}$ \citep{bola13}.

The correlator was used in the frequency domain mode with a bandwidth of 1875 MHz (488.28~kHz $\times$ 3840 channels). 
Four spectral windows were used, providing a total bandwidth of 7.5 GHz.
Bandpass, phase, and flux calibrations were done with nearby quasars. 
The second level of Quality Assurance (QA2) performed by the ALMA data reduction team was ``PASS'' except for the data set of the GRB~031203 host, where the QA2 was ``SEMIPASS'' because the synthesized beam size achieved was significantly smaller than requested and the rms noise level was higher than requested.

The data were reduced with Common Astronomy Software Applications \citep[CASA;][]{mcmu07}. 
Data calibration was done with the ALMA Science Pipeline Software of CASA versions 4.5.3, 4.7.0, and 4.7.2. 
The maps were processed with a \verb|tclean| task with the natural weighting. 
The continuum maps were created with a total bandwidth of $\sim$7.5~GHz, excluding channels with emission lines. 
Clean boxes were placed when a component with a peak signal-to-noise ratio (S/N) above 5 was identified, and {\verb CLEAN }ed down to a $2\sigma$ level. 
The GRB hosts were observed with different array configurations. 
All targets with a synthesized beam size smaller than $0\farcs7$ were {\it uv}-tapered with a value of 0\farcs8.
When a source was spatially resolved, the total flux was measured with a CASA task \verb|imfit| as an integrated flux density, otherwise the peak flux was adopted.

We detected CO emission in eight hosts at $z = 0.3$--2 (GRBs 050826, 051006, 051117B, 060814, 070306, 081109, 110918A, and 140301A). 
Because six GRB hosts were detected in CO emission so far, this study more than doubled the sample size of hosts with CO detection. 
The CO(3--2) and (4--3) line luminosities of the detected hosts were 
$L'_{\rm CO}$(3--2) $= 0.3$--$6 \times 10^9$ (K km s$^{-1}$~pc$^2$) 
and $L'_{\rm CO}$(4--3) $= 1$--$3 \times 10^9$ (K km s$^{-1}$~pc$^2$), respectively. 
The CO spectra, velocity-integrated CO maps, and continuum maps are shown in Figure~\ref{fig:results}
along with optical images taken from the Hubble Legacy Archive\footnote{\url{https://hla.stsci.edu/}} and the public data of the Dark Energy Survey \citep[DES;][]{abbo18, morg18, flau15}. 
Dust continuum emission is detected in only three hosts (GRB~051006, 051117B, and 110918A), whereas the upper limits for the nondetections were consistent with those expected from their SFRs (see Section~\ref{sec:individual}).

The CO line widths of the hosts ranged from 60 to 300 km~s$^{-1}$. 
We show the intensity-weighted velocity field maps and velocity dispersion maps for the CO-detected hosts with high significance (S/N $>$ 5.5) in Figures~\ref{fig:velocity} and \ref{fig:dispersion}, respectively. 
Two GRB hosts (GRBs 110918A and 140301A) show velocity gradient consistent with rotation with 
a line full width at half maximum (FWHM) of 200--300 km s$^{-1}$, whereas other hosts show a more disturbed velocity field.

\begin{figure*}[t]
\centering
\includegraphics[width=\linewidth]{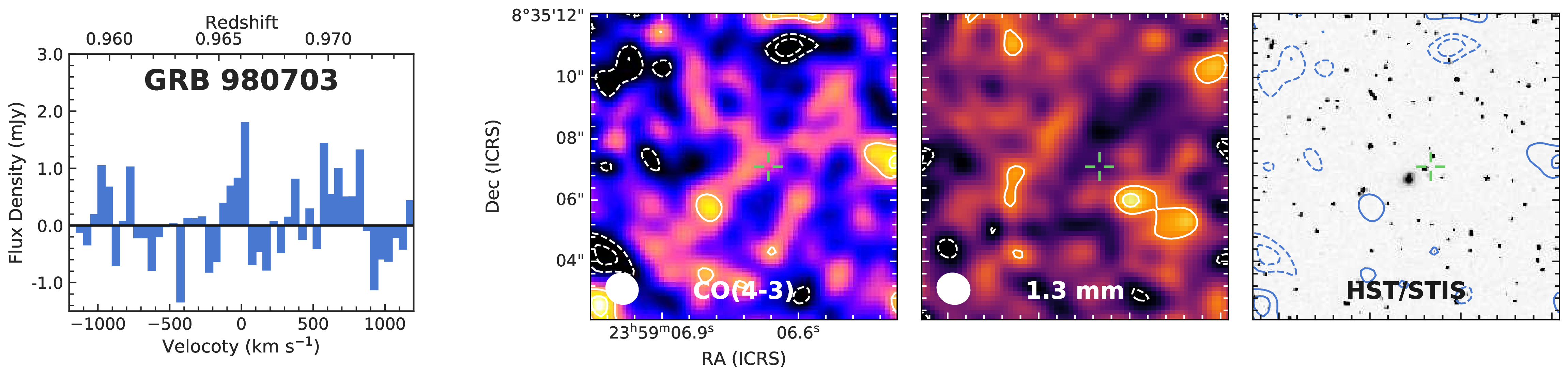}
\includegraphics[width=\linewidth]{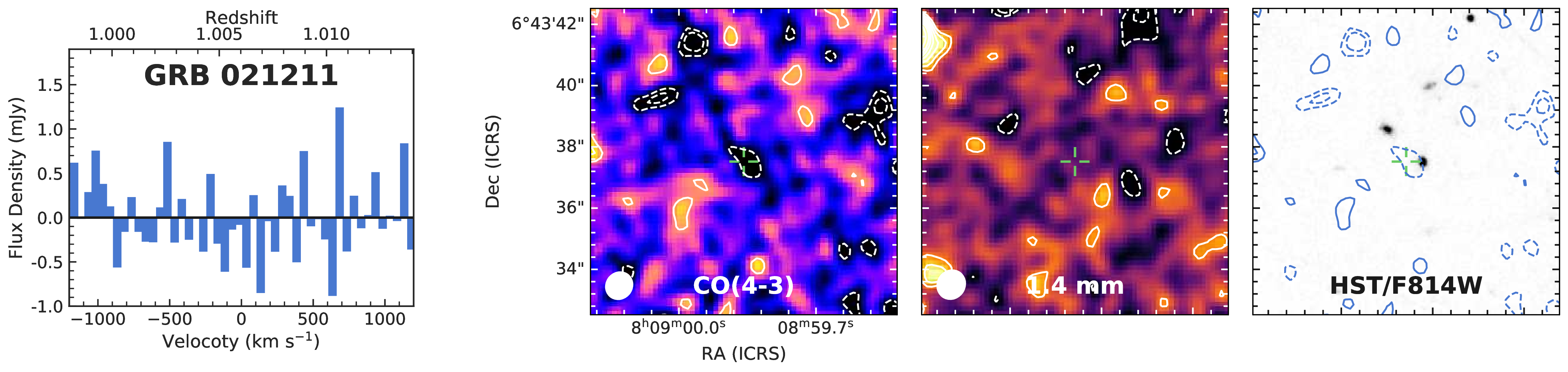}
\includegraphics[width=\linewidth]{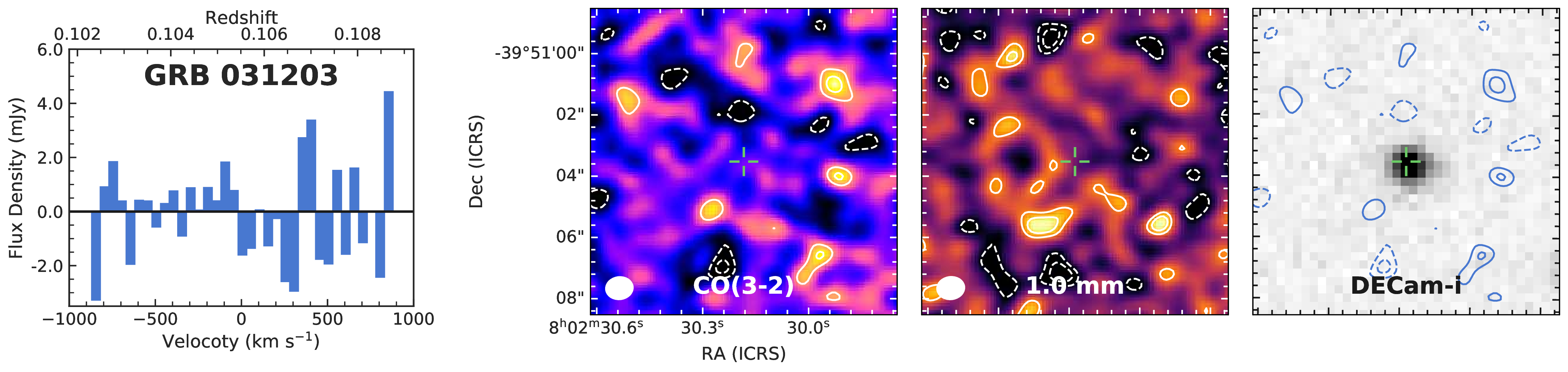}
\includegraphics[width=\linewidth]{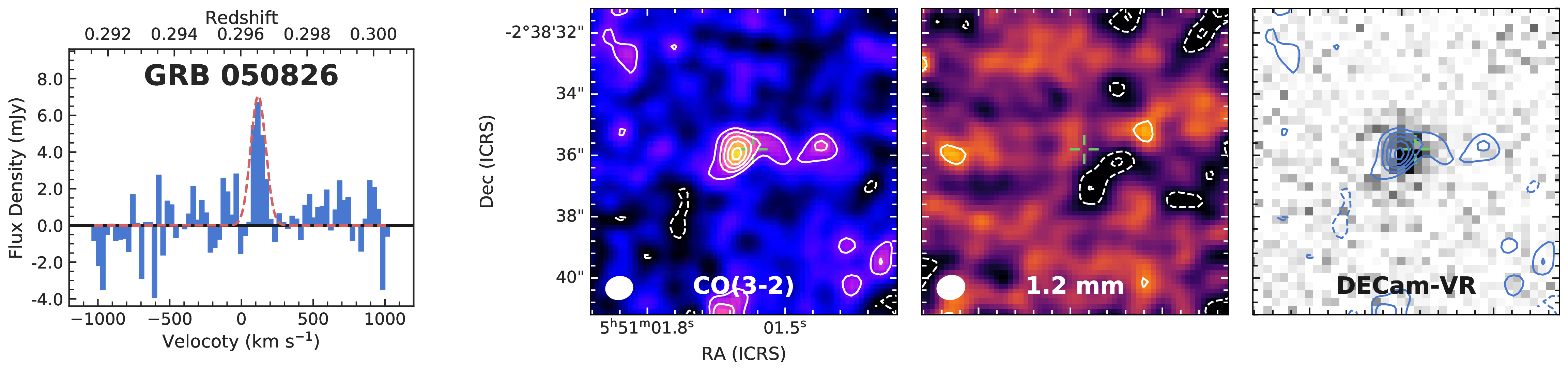}
\includegraphics[width=\linewidth]{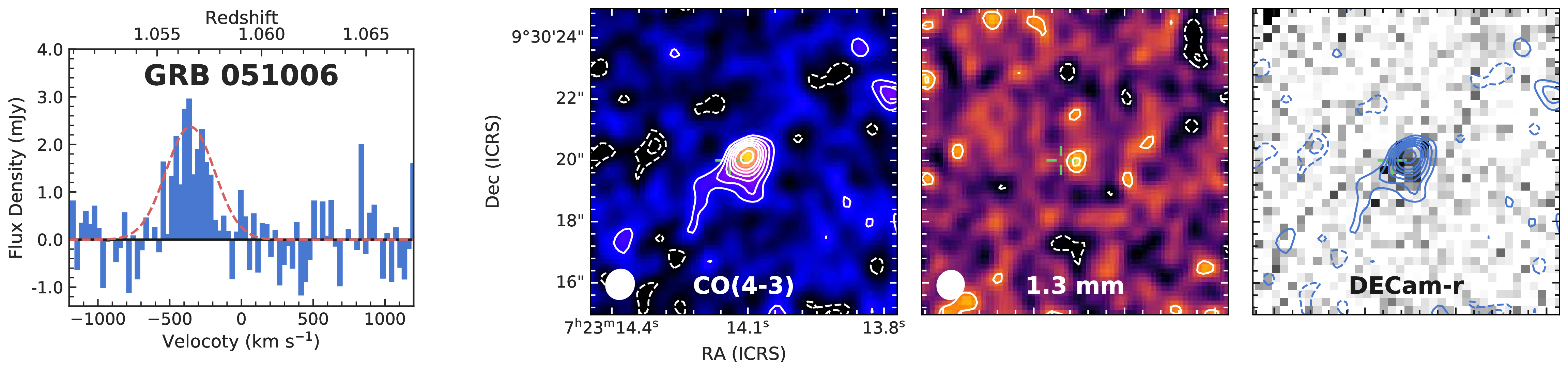}
\caption{
From left to right: CO spectra, CO velocity-integrated maps, continuum maps, and optical images. 
The image size is $10'' \times 10''$, centered at the host galaxies. 
GRB positions are marked as cross-hairs. 
The velocity resolution of the CO spectra is 30 km~s$^{-1}$ for the hosts at $\ge$$3.5\sigma$ detection and 50 km~s$^{-1}$ for the rest. 
Continuum emission is subtracted. 
The red lines show best-fitting Gaussian profiles. 
CO maps are created by integrating the channels with CO emission for the hosts at $\ge$$3.5\sigma$ detection, and the channels from $-90$ to $+90$ km~s$^{-1}$ for the rest. 
The synthesized beam size is shown in the lower left corners. 
The contours are $-3, -2, 2, 3, 5, 10$, and $15\sigma$. 
The optical images are overlaid by the contours of the CO maps.
}
\label{fig:results}
\end{figure*}

\setcounter{figure}{1}
\begin{figure*}
\centering
\includegraphics[width=\linewidth]{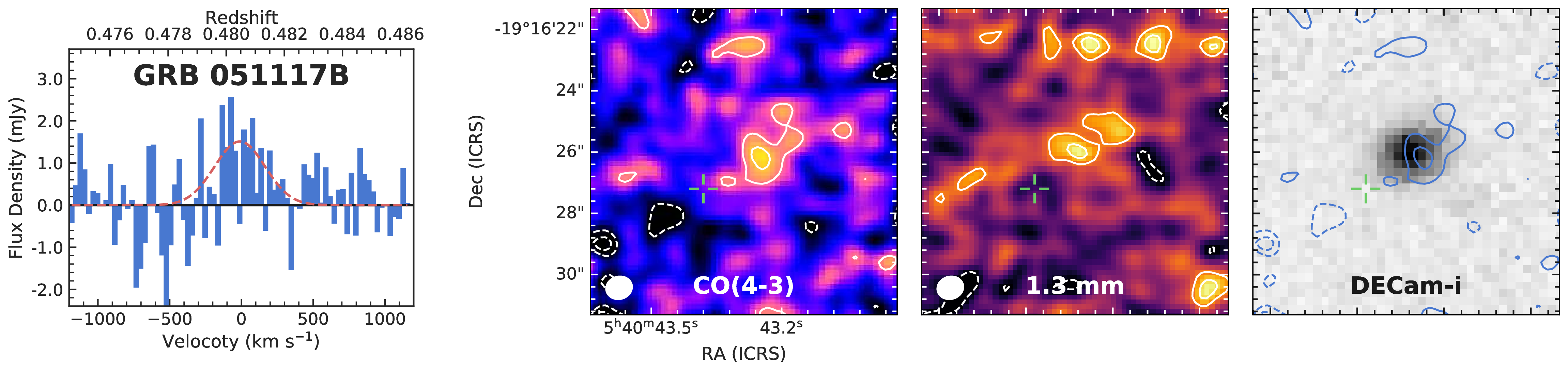}
\includegraphics[width=\linewidth]{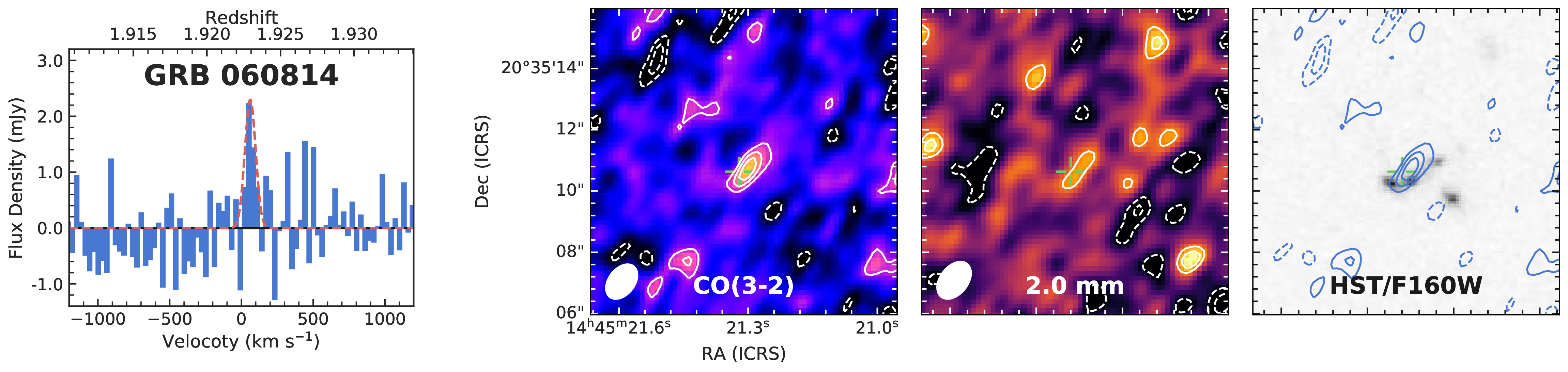}
\includegraphics[width=\linewidth]{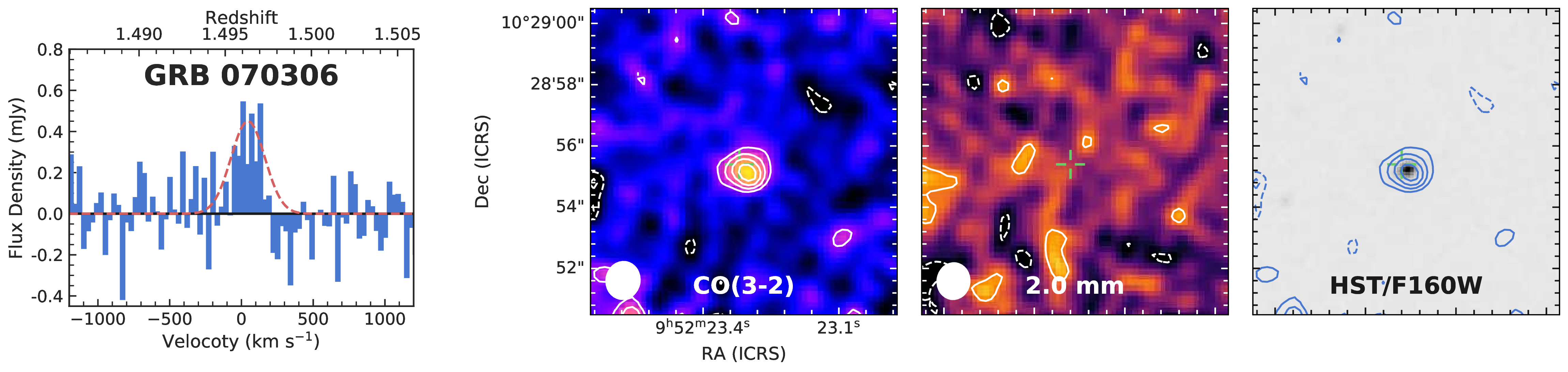}
\includegraphics[width=\linewidth]{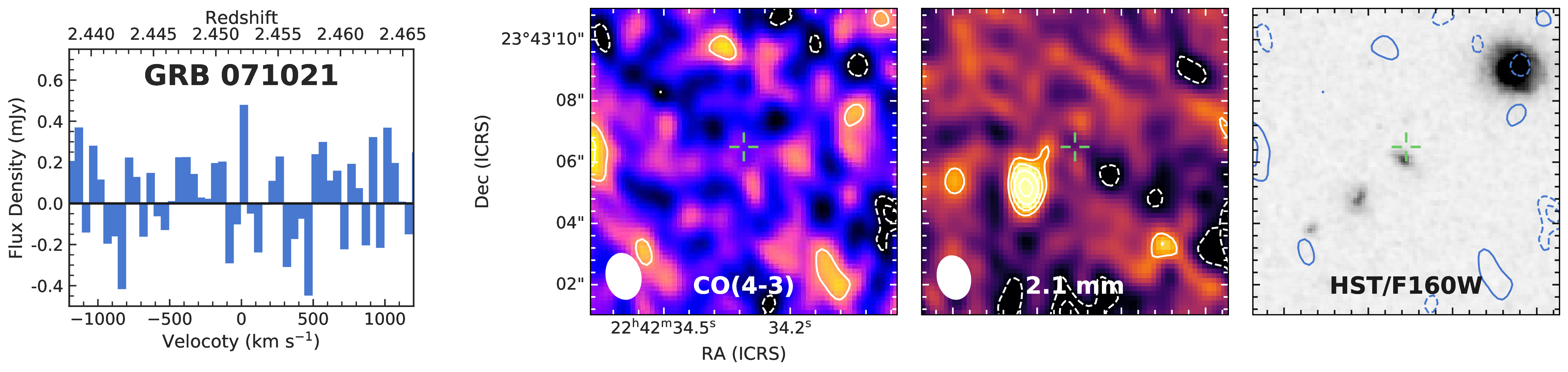}
\includegraphics[width=\linewidth]{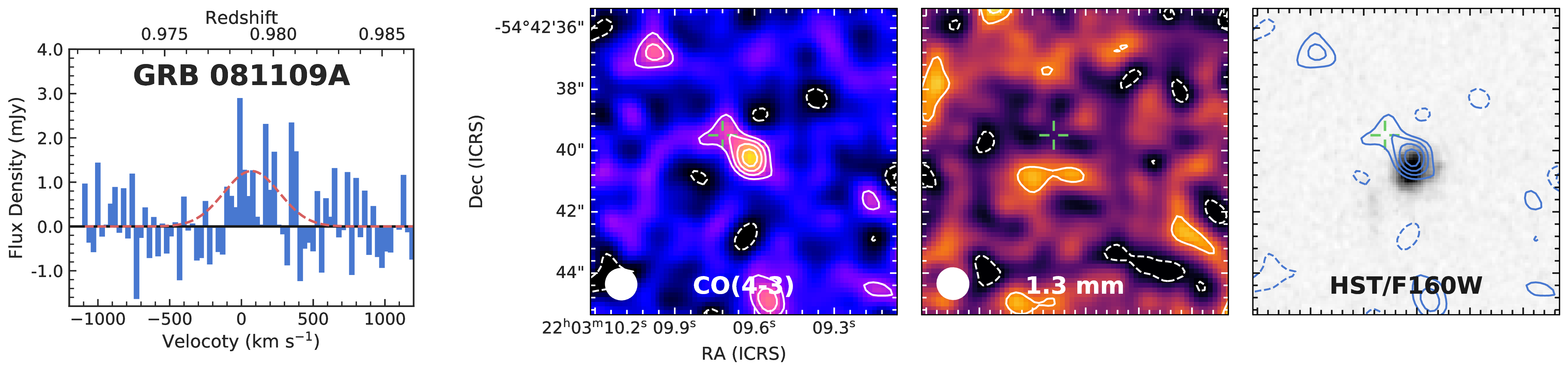}
\caption{(continued)}
\end{figure*}

\setcounter{figure}{1}
\begin{figure*}
\centering
\includegraphics[width=\linewidth]{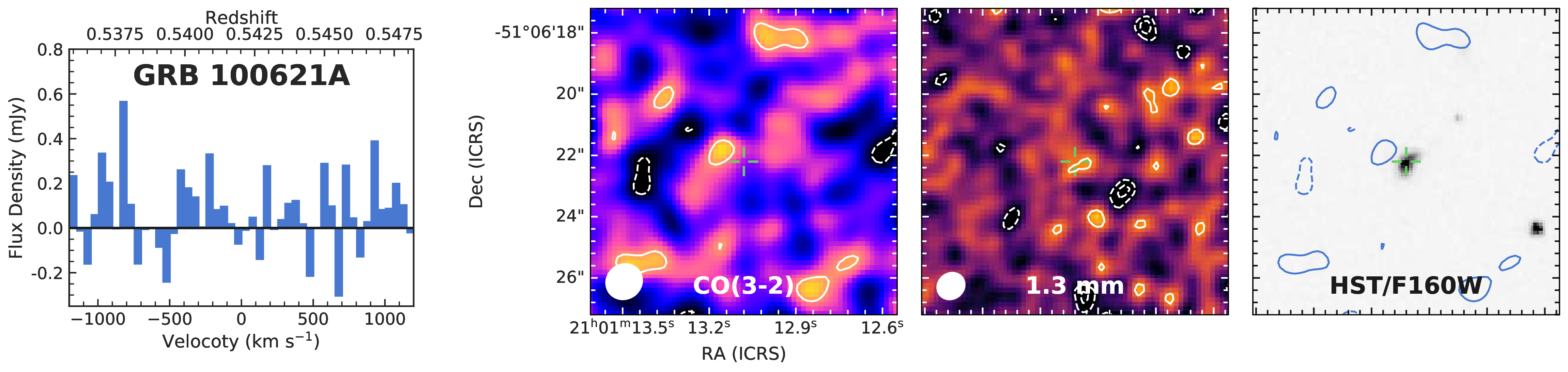}
\includegraphics[width=\linewidth]{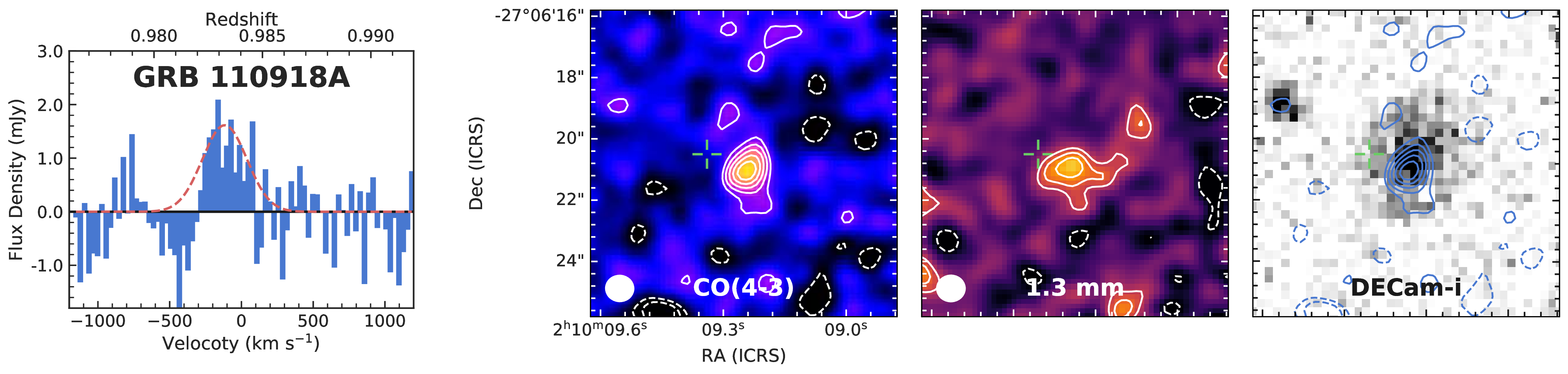}
\includegraphics[width=\linewidth]{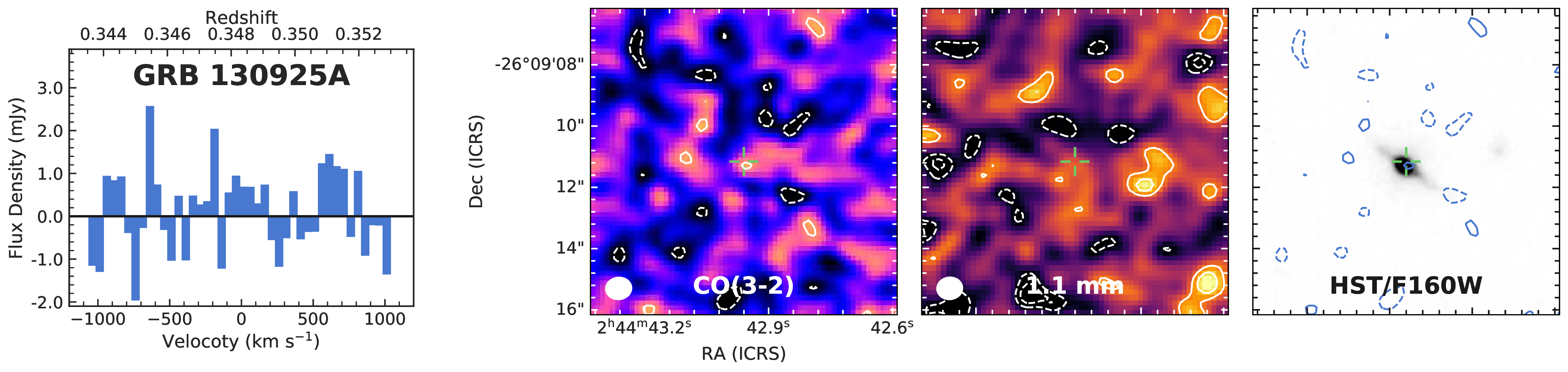}
\includegraphics[width=\linewidth]{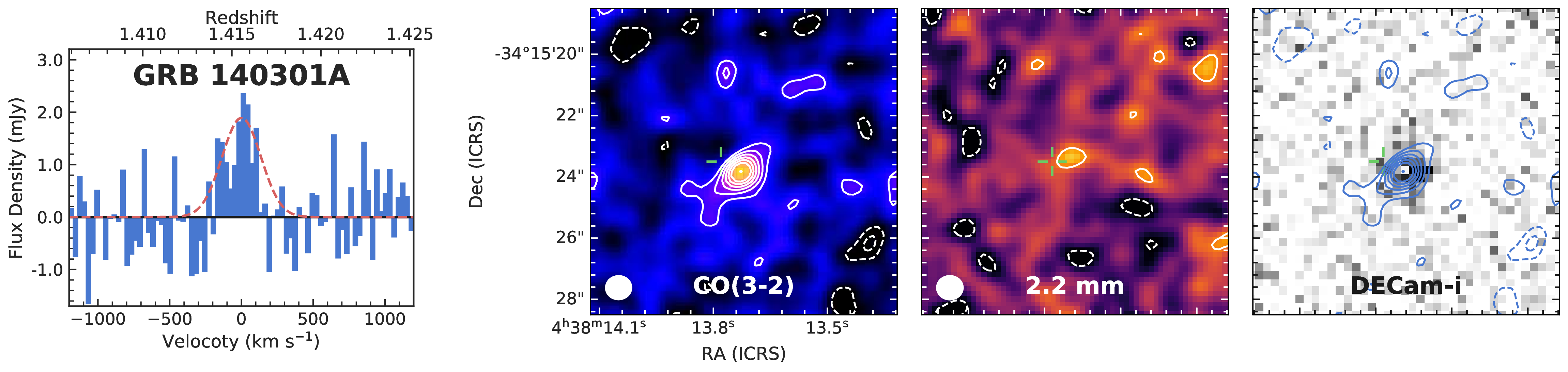}
\caption{(continued)}
\end{figure*}

\begin{figure*}
\begin{center}
\includegraphics[width=.93\linewidth]{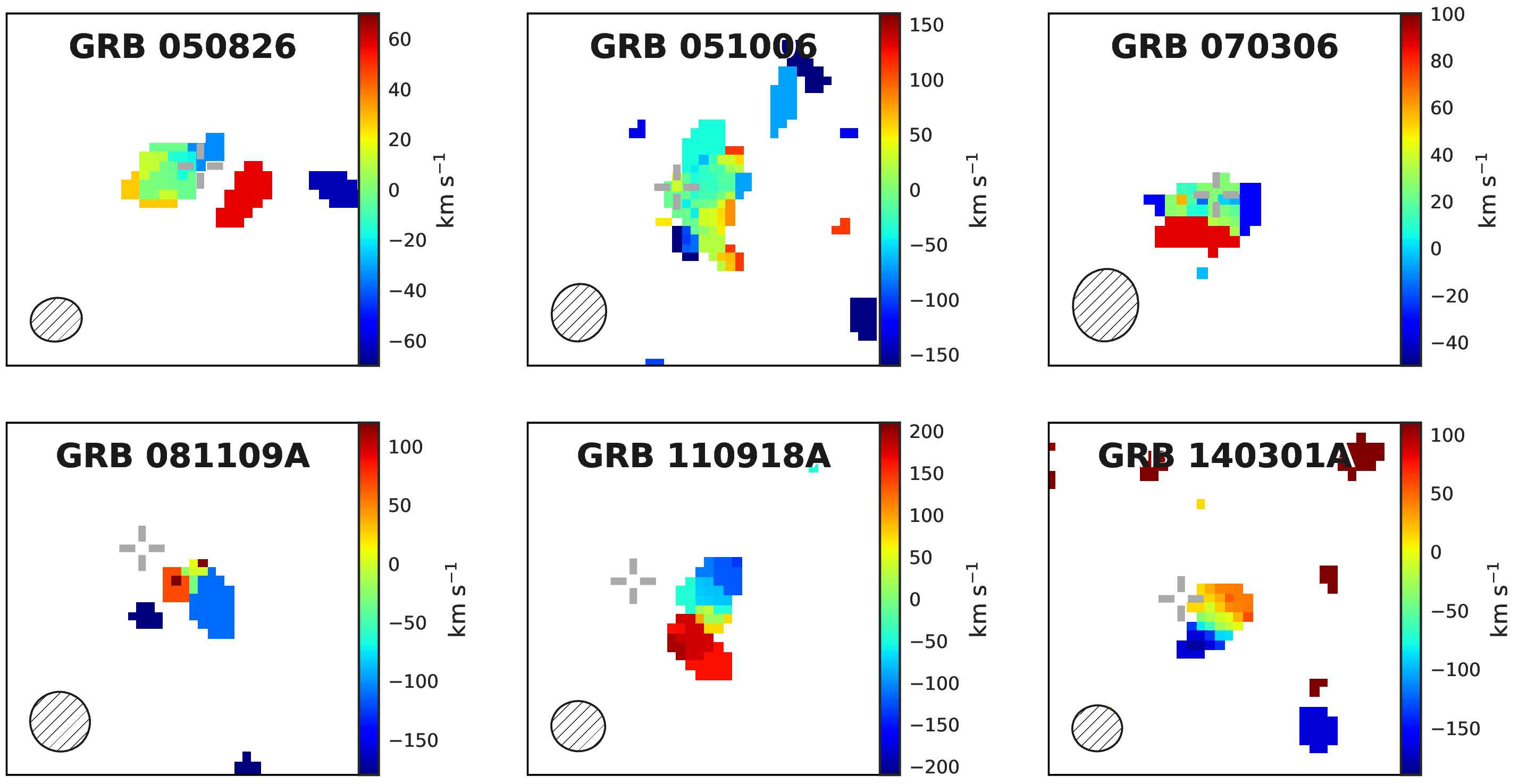}
\end{center}
\caption{
CO line intensity-weighted velocity maps of the GRB hosts with $>$5.5$\sigma$ detection. 
The emissions with $<$3$\sigma$ are clipped. 
The maps are centered at the host galaxies, and the GRB positions are marked as cross-hairs. 
The reference velocity is set to the peak velocity of Gaussian fit to the spectra. 
The synthesized beam size is shown in the lower left corners. 
The map size is $6'' \times 6''$. 
}
\label{fig:velocity}
\end{figure*}

\begin{figure*}
\begin{center}
\includegraphics[width=.93\linewidth]{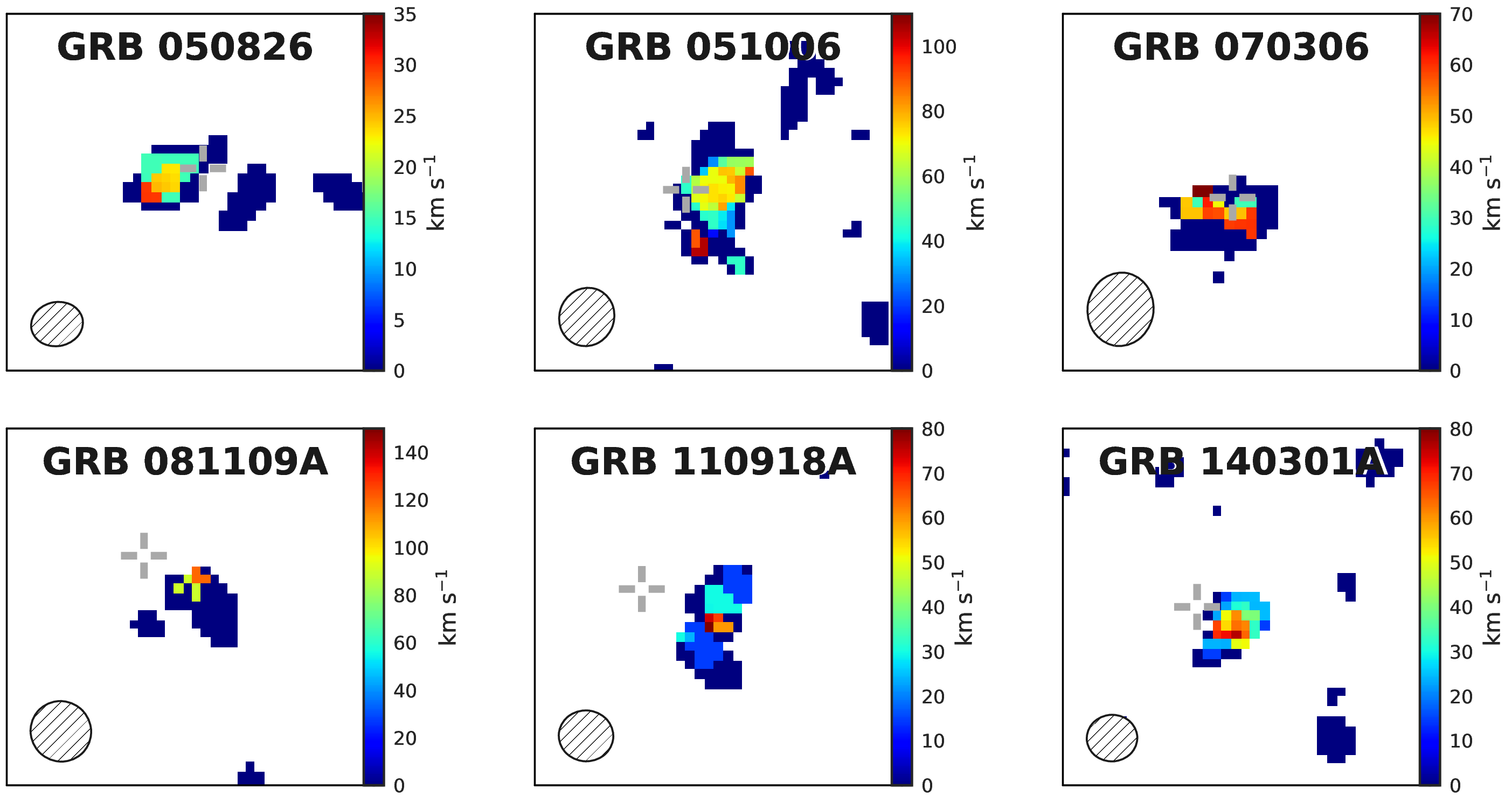}
\end{center}
\caption{
CO line intensity-weighted dispersion maps of the GRB hosts with $>$5.5$\sigma$ detection. 
The emissions with $<$3$\sigma$ are clipped. 
The maps are centered at the host galaxies, and the GRB positions are marked as cross-hairs. 
The synthesized beam size is shown in the lower left corners. 
The map size is $6'' \times 6''$. 
}
\label{fig:dispersion}
\end{figure*}

\section{SED Fit} \label{sec:sed}

In order to estimate the SFR and stellar mass of the targets in a common way, we conduct SED analysis with available photometry from UV to radio including our ALMA photometry. 
We adopt a SED modeling code of Code Investigating GALaxy Emission \citep[{\tt\string CIGALE}\footnote{\url{https://cigale.lam.fr/}};][]{burg05, noll09, boqu19}. 
{\tt\string CIGALE} is based on an energy balance principle, where the energy absorbed by dust in the UV--near-IR range is re-emitted self-consistently in the mid- and far-IR range. 
{\tt\string CIGALE} builds spectral models by computing star formation histories (SFH), stellar spectrum from the SFH and single stellar population (SSP) models, nebular emission, attenuation of the stellar and nebular emission assuming an attenuation law, and dust emission in the mid- and far-IR. 
The models are fitted to the data and physical properties are estimated through the analysis of the likelihood distribution.

We adopt a SFH of the delayed SFH with an optional exponential burst ({\tt\string sfhdelayed}) with a form of SFR $\propto t/\tau^2 \exp(-t/\tau)$, where $\tau$ is the time at which the SFR peaks. 
The SSP library of \cite{bruz03} ({\tt\string bc03}) and the \cite{chab03} IMF are adopted for computing the spectrum of composite stellar populations. 
The nebular emission is modeled based on \cite{inou11}, where {\tt\string CLOUDY} \citep{ferl98, ferl13} is used. 
The attenuation by dust is calculated with the attenuation law of {\tt\string dustatt\_modified\_starburst}, which is based on the \cite{calz00} starburst attenuation curve. 
The dust emission is computed with the module of {\tt\string dl2014}, which is based on the model of \cite{drai07} and \cite{drai14}. 
The synchrotron radio emission is calculated for the GRB hosts with radio detection (GRBs 980703, 031203, 051006, and 060814) with the {\tt\string radio} module, which relies on the radio-IR correlation of \cite{helo85}.

The photometry data of our targets are taken from the GHostS database (\url{www.grbhosts.org} and references therein), \cite{perl16}, and \cite{perl17a}. 
The data for the GRB~140301A host is taken from the database of the Dark Energy Survey. 
We also use the photometry and upper limits obtained in our ALMA observations.

\begin{figure*}
\begin{center}
\includegraphics[width=\linewidth]{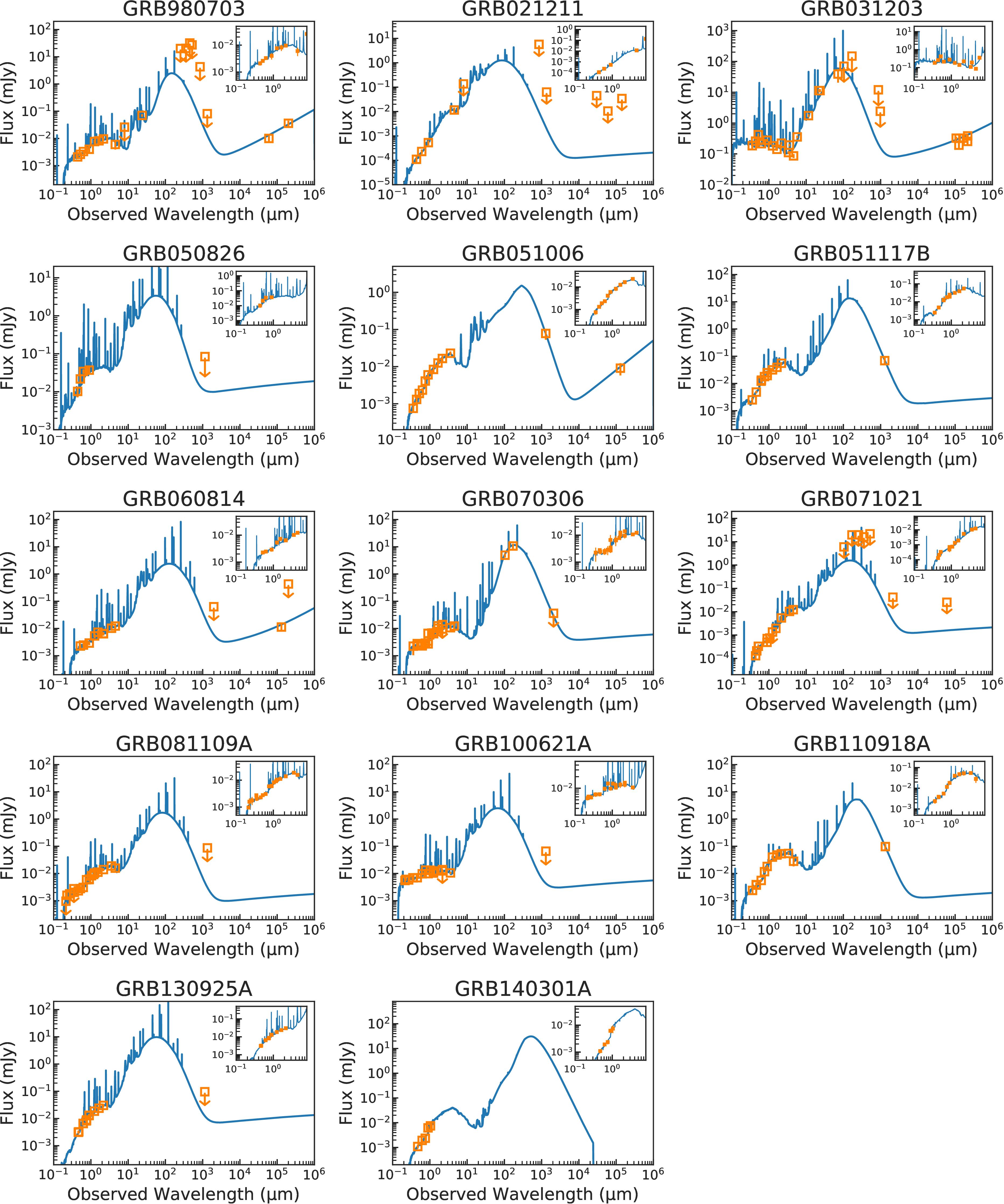}
\end{center}
\caption{
Best-fit SED models with photometry data (squares). 
Arrows represent upper limits. 
}
\label{fig:sed}
\end{figure*}

The results on SFR and stellar mass are presented in Table~\ref{tab:sed} and the SEDs are shown in Figure~\ref{fig:sed}. 
The stellar masses are overall consistent with previous studies presented in Table~\ref{tab:targets}. 
We adopt the SFRs and stellar masses for the GRB hosts derived in this study in subsequent discussions.

\begin{deluxetable}{ccc}
\tablecaption{Results of SED Fit.\label{tab:sed}}
\tablewidth{0pt}
\tablehead{
\colhead{GRB} & \colhead{SFR} & \colhead{$M_*$} \\ 
\colhead{} & \colhead{($M_{\odot}$~yr$^{-1}$)} & \colhead{($M_{\odot}$)} 
}
\startdata
 980703 & $14 \pm  3$ & $(7.9 \pm 2.1) \times 10^{ 9}$ \\ 
 021211 & $ 9.9 \pm  6.0$ & $(3.3 \pm 2.6) \times 10^{ 9}$ \\ 
 031203 & $ 2.9 \pm  0.6$ & $(3.3 \pm 0.5) \times 10^{ 8}$ \\ 
 050826 & $ 5.0 \pm  4.2$ & $(4.4 \pm 2.6) \times 10^{ 9}$ \\ 
 051006 & $61 \pm 18$ & $(2.6 \pm 0.6) \times 10^{10}$ \\ 
051117B & $14 \pm 12$ & $(1.2 \pm 0.2) \times 10^{10}$ \\ 
 060814 & $67 \pm 10$ & $(1.2 \pm 0.3) \times 10^{10}$ \\ 
 070306 & $38 \pm 28$ & $(1.2 \pm 0.3) \times 10^{10}$ \\ 
 071021 & $40 \pm 29$ & $(8.8 \pm 3.3) \times 10^{10}$ \\ 
081109A & $21 \pm  9$ & $(1.3 \pm 0.3) \times 10^{10}$ \\ 
100621A & $ 9.2 \pm  1.3$ & $(9.9 \pm 1.2) \times 10^{ 8}$ \\ 
110918A & $22 \pm  7$ & $(6.7 \pm 1.3) \times 10^{10}$ \\ 
130925A & $ 3.9 \pm  1.2$ & $(1.6 \pm 0.6) \times 10^{ 9}$ \\ 
140301A & $ 233 \pm  138$ & $(4.4 \pm 2.5) \times 10^{10}$ \\ 
\enddata
\end{deluxetable}

\section{Properties of the GRB Host Galaxies} \label{sec:properties}

We derived molecular gas mass from the CO emission, dust mass, and far-infrared (FIR)-based SFR from the continuum emission. 

\begin{deluxetable*}{cccccccccc}[t]
\tabletypesize{\scriptsize}
\tablecaption{Results on CO line.\label{tab:results_co}}
\tablewidth{0pt}
\tablehead{
\colhead{GRB} & \colhead{$z_{\rm CO}$$^a$} & \colhead{$\Delta v$$^b$} & \colhead{$S\Delta v$$^c$} & 
\colhead{S/N$^d$} & \colhead{$L'_{\rm CO}$$^e$} & \colhead{$M_{\rm gas}$$^f$} & 
\colhead{$\alpha_{\rm CO}(Z)$$^g$} &  \colhead{$\mu_{\rm gas}$$^h$} & \colhead{$t_{\rm depl}$$^i$} \\
\colhead{} & \colhead{} & \colhead{(km s$^{-1}$)} & \colhead{(Jy km s$^{-1}$)} & \colhead{} & 
\colhead{(K km s$^{-1}$~pc$^2$)} & \colhead{($M_{\odot}$)} & 
\colhead{$M_{\odot}$~(K~km~s$^{-1}$~pc$^2$)$^{-1}$} & \colhead{} & \colhead{(Gyr)} 
}
\startdata
980703 &--   &--        &$<$0.13      & -- &$<$$4.0   \times10^8$&$<$$1.1\times10^{10}$     &12  &$<$1.4       &$<$0.76       \\
021211 &--   &--        &$<$0.08      & -- &$<$$2.9   \times10^8$&$<$$6.6\times10^{ 9}$     & 9.6&$<$2.0       &$<$0.66       \\
031203 &--   &--        &$<$3.0       & -- &$<$$1.8   \times10^8$&$<$$3.5\times10^{ 9}$     &14  &$<$11        &$<$1.2        \\
050826 &0.296&$ 81\pm17$&$0.57\pm0.09$& 6.5&$2.8\pm0.5\times10^8$&$(2.9\pm0.5)\times10^{ 9}$& 6.2&$0.67\pm0.41$&$0.59\pm0.50$ \\
051006 &1.057&$263\pm39$&$0.69\pm0.07$& 10 &$2.7\pm0.3\times10^9$&$(3.6\pm0.4)\times10^{10}$& 5.5&$1.4 \pm0.4$ &$0.59\pm0.18$ \\
051117B&0.480&$273\pm42$&$0.66\pm0.19$& 3.5&$9.1\pm0.3\times10^8$&$(5.9\pm1.7)\times10^{ 9}$& 3.9&$0.47\pm0.16$&$0.44\pm0.39$ \\
060814 &1.923&$ 58\pm19$&$0.22\pm0.04$& 5.2&$2.7\pm0.5\times10^9$&$(5.7\pm1.1)\times10^{10}$&14  &$4.6\pm1.5$  &$0.85\pm0.21$ \\
070306 &1.496&$183\pm42$&$0.17\pm0.03$& 5.7&$2.2\pm0.4\times10^9$&$(2.9\pm0.5)\times10^{10}$&11  &$2.4\pm0.7$  &$0.76\pm0.59$ \\
071021 &--   &--        &$<$0.07      & -- &$<$$1.3   \times10^9$&$<$$4.6\times10^{10}$     &15  &$<$0.52      &$<$1.2        \\
081109A&0.979&$302\pm46$&$0.42\pm0.07$& 5.8&$1.4\pm0.2\times10^9$&$(2.0\pm0.3)\times10^{10}$& 5.8&$1.5 \pm0.4 $&$0.96\pm0.46$ \\
100621A&--   &--        &$<$0.08      & -- &$<$$1.4   \times10^8$&$<$$1.9\times10^{ 9}$     & 8.3&$<$1.9       &$<$0.20       \\
110918A&0.983&$242\pm62$&$0.52\pm0.07$& 6.9&$1.7\pm0.2\times10^9$&$(1.9\pm0.3)\times10^{10}$& 4.4&$0.28\pm0.07$&$0.84\pm0.30$ \\
130925A&--   &--        &$<$0.11      & -- &$<$$7.6   \times10^7$&$<$$7.6\times10^{ 8}$     & 6.0&$<$0.48      &$<$0.20       \\
140301A&1.416&$211\pm46$&$0.46\pm0.05$& 9.1&$5.5\pm0.6\times10^9$&$(4.3\pm0.5)\times10^{10}$& 4.7&$0.99\pm0.56$&$0.17\pm0.09$ \\
\enddata
\tablecomments{
Limits are 3$\sigma$. 
$^a$ Redshift derived from the CO line. 
$^b$ CO line width (FWHM). 
$^c$ Velocity-integrated CO flux. Assuming a velocity width of 180 km s$^{-1}$ for non detection. 
$^d$ Peak signal-to-noise ratio in the velocity-integrated map. 
$^e$ Line luminosity of CO(3--2) or CO(4--3). 
$^f$ Molecular gas mass. 
$^g$ Metallicity-dependent CO-to-H$_2$ conversion factor used in this study. 
$^h$ Molecular gas mass fraction ($M_{\rm gas}/M_*)$. 
$^i$ Molecular gas depletion timescale ($M_{\rm gas}/{\rm SFR})$. 
}
\end{deluxetable*}

\subsection{CO Line Luminosity Ratios} \label{sec:ratio}

The CO luminosity of a ground rotational transition $J = 1$--0 is required for deriving the molecular gas mass by applying the conversion factor, and we need to assume a CO line ratio in the case of host galaxies with only higher $J$ CO lines. 
The ratios of CO line luminosities in GRB hosts have been unexplored because of the limited number of CO observations and the lack of detection of multiple CO transitions. 
\cite{hats19} reported the line ratios for the GRB~080207 host with upper $J$ levels from 1 to 4. 
They found that the line ratios were close to unity up to $J = 4$, similar to those of local starburst M~82, local (U)LIRGs, and QSOs/radio galaxies, suggesting a high molecular gas excitation condition in the host.

Among the samples in this study, the hosts of GRB~031203 and GRB~060814 were observed in other CO transitions in previous studies. 
\cite{mich18} reported upper limits on the CO(2--1) line in the hosts of GRBs 031203 and 060814. 
We detected the CO(3--2) line in the GRB~060814 host, allowing us to constrain the line ratio. 
The obtained lower limit for the line luminosity ratio was CO(3--2)/CO(2--1) $> 0.32$ (3$\sigma$). 
This is consistent with the line ratio for the GRB~080207 host, although the constraint was weak. 
Because the CO line ratio for GRB hosts was obtained only in the GRB~080207 host, in this study we adopted the line ratios of CO(3--2)/CO(1--0) $=r_{31}$ $=0.6$ and CO(4--3)/CO(1--0) $=r_{41}$ $=0.4$, which are the intermediate values between Milky Way and M~82 \citep{cari13} and appropriate for $z \sim 1$--2 MS galaxies \citep{dann09, arav14, dadd15}. 
We note that the derived molecular gas mass could vary by a factor of a few due to the uncertainty of CO line ratios.

\subsection{Molecular Gas Mass} \label{sec:gas}

The CO luminosity was calculated as follows \citep{solo05}
\begin{eqnarray}
L'_{\rm CO} = 3.25 \times 10^7 S_{\rm CO}\Delta v \nu_{\rm obs}^{-2} D_{\rm L}^2 (1+z)^{-3}, 
\end{eqnarray}
where $L'_{\rm CO}$ is in K km s$^{-1}$~pc$^2$, $S_{\rm CO}\Delta v$ is the velocity-integrated intensity in Jy~km~s$^{-1}$, $\nu_{\rm obs}$ is the observed line frequency in GHz, and $D_{\rm L}$ is the luminosity distance in Mpc.
The molecular gas mass is derived from 
\begin{eqnarray}
M_{\rm gas} = \alpha_{\rm CO} L'_{\rm CO(1-0)}, 
\end{eqnarray}
where $\alpha_{\rm CO}$ is the CO-to-H$_2$ conversion factor, including the contribution of helium mass. 
$L'_{\rm CO(1-0)}$ is calculated from the observed $L'_{\rm CO(3-2)}$ or $L'_{\rm CO(4-3)}$ devided by $r_{31}$ or $r_{41}$ (Section \ref{sec:ratio}).
The conversion factor is thought to be dependent on gas-phase metallicity, increasing $\alpha_{\rm CO}$ with decreasing metallicity \citep[e.g.,][]{wils95, arim96, kenn12, bola13}. 
We adopted the relation between metallicity and $\alpha_{\rm CO}$ of \cite{genz15}, where they took the geometric mean of the empirical relations of \cite{genz12} and \cite{bola13} and derived a relation for the local and high-redshift sample. 
To apply the relation, we converted the metallicity to the calibration of \cite{pett04} by using the metallicity conversion of \cite{kewl08}.
We derived metallicity from the mass--metallicity conversion of \cite{genz15} for the hosts where metallicity was not obtained from emission line diagnostics. 
The adopted metallicity-dependent conversion factor $\alpha_{\rm CO}(Z)$ is 4--15 $M_{\odot}$~(K~km~s$^{-1}$~pc$^2$)$^{-1}$ (Table~\ref{tab:results_co}). 
Because the approach of \cite{genz15} is not applicable for significantly sub-solar metallicity galaxies ($12+\log({\rm O/H}) \lesssim 8.4$), where the relations of \cite{genz12} and \cite{bola13} deviate from each other, we adopted the harmonic mean of the two relations by following \cite{tacc18}. 
If we had adopted a Galactic conversion factor of $\alpha_{\rm CO} = 4.4$ $M_{\odot}$~(K~km~s$^{-1}$~pc$^2$)$^{-1}$ \citep{bola13} regardless of metallicity, which is appropriate for $z \sim 1$--2 normal star-forming galaxies \citep{dadd10, tacc13}, derived molecular gas mass would be smaller at most by a factor of 3 for hosts with a larger $\alpha_{\rm CO}(Z)$. 
Note that although $\alpha_{\rm CO} = 0.8$ $M_{\odot}$~(K~km~s$^{-1}$~pc$^2$)$^{-1}$ is derived for ultra-luminous IR galaxies (ULIRGs) (galaxies with $L_{\rm FIR} > 10^{12}$~$L_{\odot}$) \citep{down98}, our targets are not categorized as ULIRGs (Section~\ref{sec:individual}).

For non-detection hosts, we derived 3$\sigma$ upper limits by assuming a line width of 180~km~s$^{-1}$, which is a median value for $z \sim 1$ star-forming galaxies derived from CO observations \citep{tacc13} and is also comparable to the mean velocity width (200~km~s$^{-1}$) of the CO-detected GRB hosts in this study.

\begin{figure}
\begin{center}
\includegraphics[width=\linewidth]{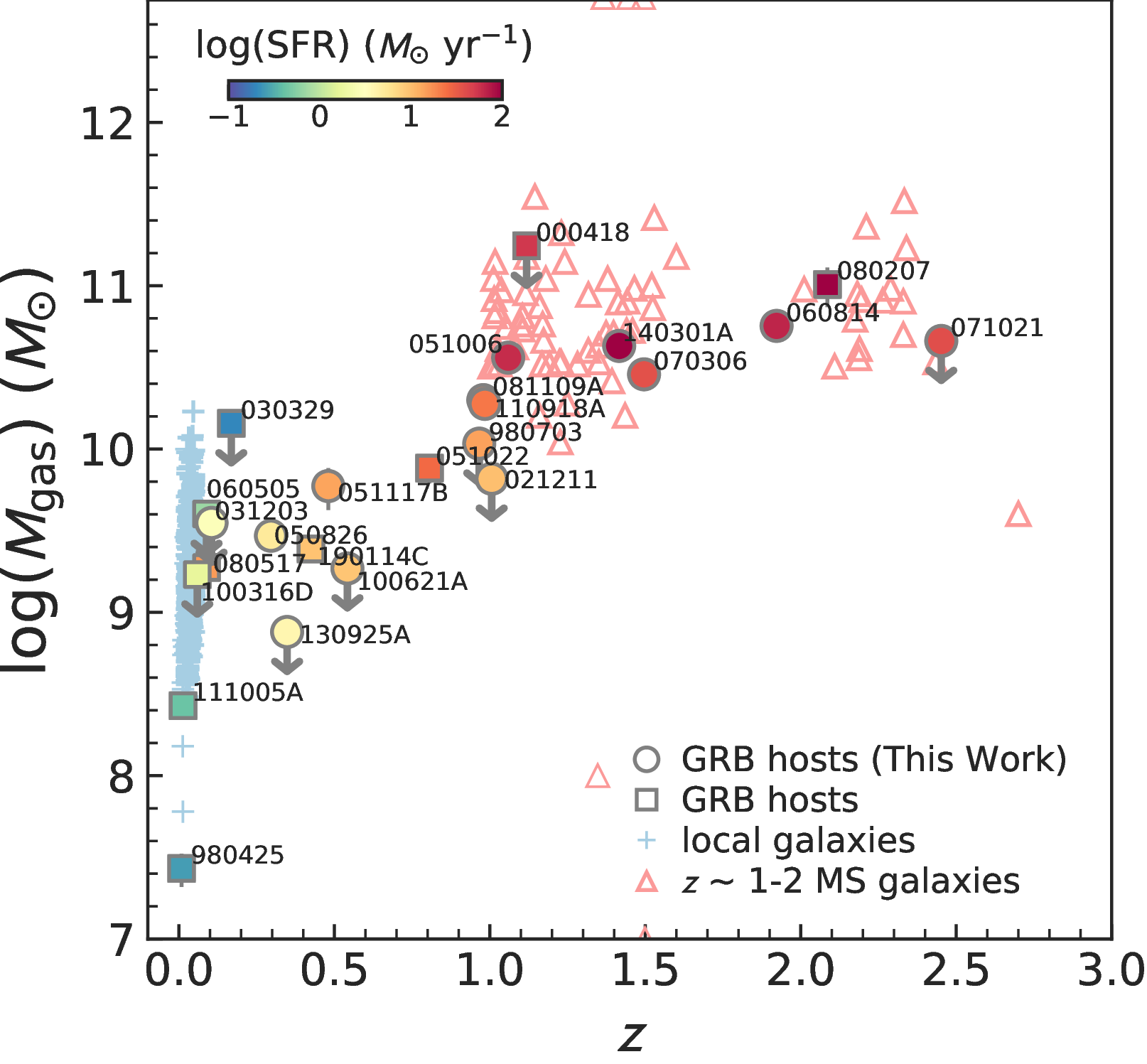}
\end{center}
\caption{
Molecular gas mass as a function of redshift. 
Circles and squares represent GRB hosts of our targets and in the literature, respectively. 
Arrows represent 3$\sigma$ upper limits. 
For comparison, we plot $z \sim 1$--2 MS galaxies \citep{dadd10, magn12, magd12, tacc13, seko16a} and local star-forming galaxies \citep{both14, sain17}. 
The GRB hosts are color coded by SFR. 
}
\label{fig:z-mgas}
\end{figure}

\begin{figure}
\begin{center}
\includegraphics[width=\linewidth]{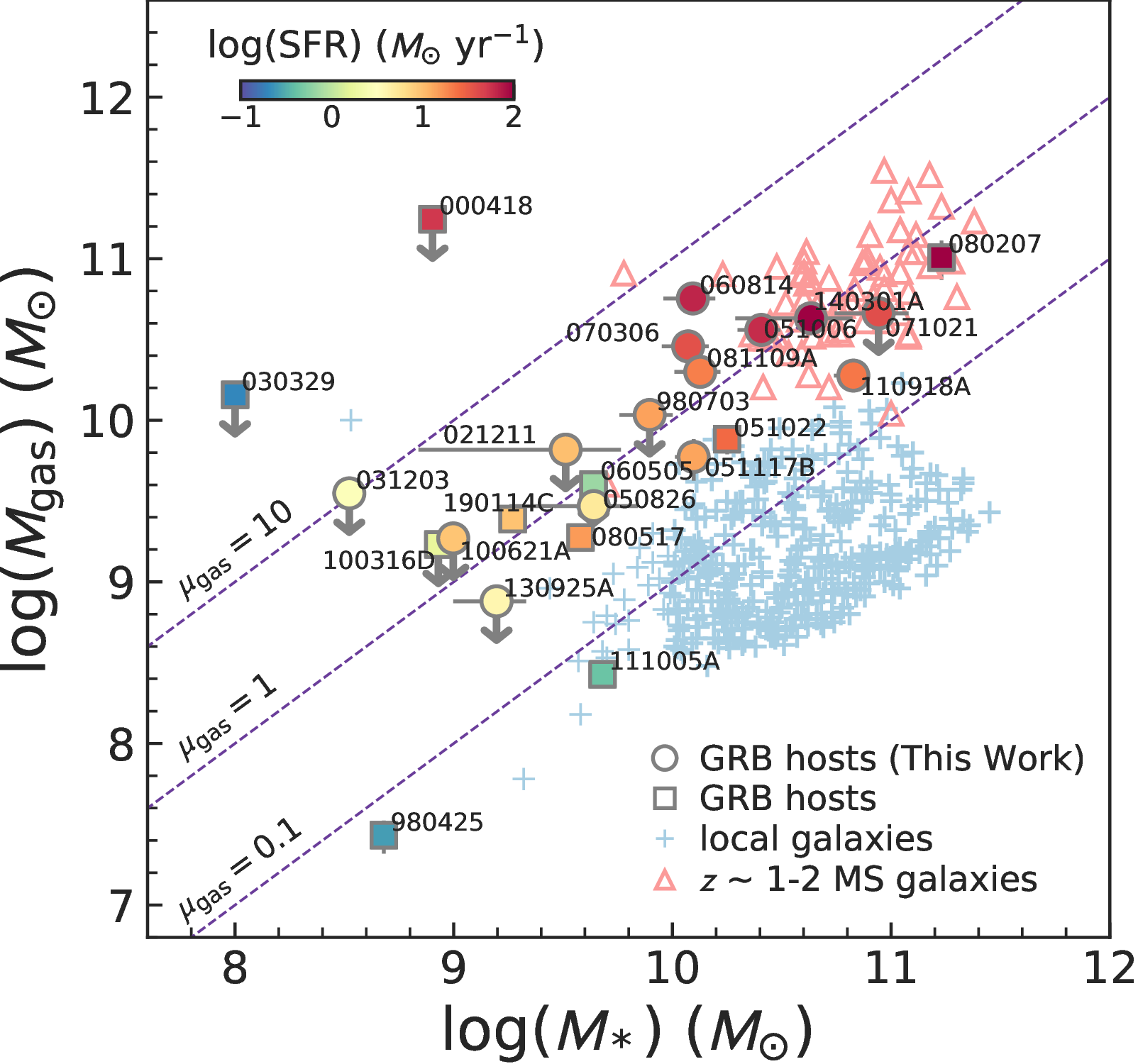}
\end{center}
\caption{
Molecular gas mass--stellar mass plot. 
Symbols are the same as in Figure~\ref{fig:z-mgas}. 
The diagonal lines represent molecular gas fractions ($\mu_{\rm gas} = M_{\rm gas}/M_*$) of 0.1, 1, and 10. 
The GRB hosts are color coded by SFR. 
}
\label{fig:mstar-mgas}
\end{figure}

The derived molecular gas mass is (0.2--$6) \times 10^{10}$~$M_{\odot}$ (Table~\ref{tab:results_co}), which is comparable to that of GRB hosts at similar redshifts reported in the literature \citep{hats14, stan15, mich18, arab18, hats19, deug19}. 
Figures~\ref{fig:z-mgas} and \ref{fig:mstar-mgas} show the molecular gas mass as a function of redshift and stellar mass, respectively. 
We show the CO-detected GRB hosts in our sample and in the literature (Table~\ref{tab:other_hosts}), $z \sim 1$--2 MS galaxies \citep{dadd10, magn12, magd12, tacc13}, and local star-forming galaxies in \cite{both14} (ALLSMOG sample at $0.01 < z < 0.03$) and \cite{sain17} (xCOLD GASS sample at $0.01 < z < 0.05$). 
Most of the CO-detected GRB hosts at $z \gtrsim 1$ show molecular gas fraction ($\mu_{\rm gas} = M_{\rm gas}/M_*$) comparable to those of $z \sim 1$--2 MS galaxies in Figure \ref{fig:mstar-mgas}. 
On the other hand, some of the hosts with $\log(M_*/M_{\odot}) \lesssim 10$ at lower redshifts have a higher gas fraction compared to local galaxies of similar stellar mass, which could indicate some excess of molecular gas.

\subsection{Dust Mass, FIR Luminosity, and SFR} \label{sec:dust}

The FIR luminosity and dust mass were derived as \citep{debr03}
\begin{eqnarray}
& & L_{\rm FIR} = 4\pi M_{\rm dust} \int_0^{\infty} \kappa_d(\nu_{\rm{rest}})B(\nu_{\rm{rest}}, T_{\rm dust}) d\nu, \\
& & M_{\rm dust} = \frac{S_{\rm{obs}}D_L^2}{(1+z)\kappa_d(\nu_{\rm{rest}})B(\nu_{\rm{rest}}, T_{\rm dust})},
\end{eqnarray}
where $\kappa_d(\nu_{\rm{rest}})$ is the dust mass absorption coefficient, $\nu_{\rm{rest}}$ is the rest-frame frequency, $T_{\rm dust}$ is the dust temperature, $B(\nu_{\rm{rest}}, T_{\rm dust})$ is the Planck function, and $S_{\rm{obs}}$ is the observed flux density.
We assumed that the absorption coefficient varies as $\kappa_d \propto \nu^{\beta}$ and that the emissivity index lies between 1 and 2 \citep[e.g.,][]{hild83}, and adopt $\kappa_d(125\ \mu m) = 2.64 \pm 0.29$~m$^2$~kg$^{-1}$ \citep{dunn03} and $\beta = 1.5$. 
SFRs were derived from SFR $= 1.72 \times 10^{-10} L_{\rm FIR}$ \citep{kenn98} and scaled to \cite{chab03} IMF. 
Dust temperatures for GRB hosts have not been well constrained. 
\cite{mich08} studied four submillimeter-detected GRB hosts and derived the dust temperature of $T_{\rm dust} = 44$--51 K, with the average value of 47.5 K. 
\cite{hunt14} derived the dust temperature of 17 GRB hosts (including the hosts of GRBs 980703 and 070306) with {\sl Herschel} observations by fitting their SEDs and the dust temperature ranged from 21 to 132~K, with an average temperature of 48~K. 
\cite{hats19} and \cite{hash19} estimated the dust temperature of 37--40 K for the GRB~080207 host. 
In this study, we adopted $T_{\rm dust} = 40$ K and derived physical quantities (3$\sigma$ limits for nondetections). 
Note that dust temperature could increase with redshift, as has been found for star-forming galaxies \citep[e.g.,][]{magd12, magn13, beth15, schr18}. 
If we adopt $T_{\rm dust} = 30$ or 50 K, the dust mass would change by a factor of 1.5 and 0.8, and SFR would change by a factor of 0.3 and 2.6. 
The dust mass and SFR for continuum-detected hosts are $\sim$$10^7$ $M_{\odot}$ and $\sim$10--80 $M_{\odot}$~yr$^{-1}$, respectively. 
The derived quantities are presented in Table~\ref{tab:results_dust}. 
No GRB hosts were found to be in the class of ULIRGs ($L_{\rm IR} > 10^{12}$ $L_{\odot}$).

The molecular gas-to-dust mass ratios for the GRB hosts with both CO and continuum detections are $M_{\rm gas}/M_{\rm dust} = 1730\pm560$, $640\pm260$, and $840\pm210$ for the hosts of GRBs 051006, 051117B, and 110918A, respectively. 
We also obtained lower limits for the hosts with CO detection only is $M_{\rm gas}/M_{\rm dust} > 600$--1000. 
It is shown that gas-to-dust ratio increase with decreasing metallicity for local and $z \sim 1$--3 star-forming galaxies \citep[e.g.,][]{lero11, magd12, sain13, remy14}. 
It is also suggested that gas-to-dust ratios for $z \sim 1$--3 star-forming galaxies are higher than those of local galaxies \citep{sain13, seko14}. 
The gas-to-dust ratios obtained for $z \sim 1.4$ MS galaxies are $M_{\rm gas}/M_{\rm dust} = 200$--1500 \citep{seko16a, seko16b}. 
By considering the dependence of gas-to-dust ratio on metallicity or redshift, the ratios for the GRB hosts are comparable to other star-forming galaxies.

\begin{deluxetable*}{ccccccc}
\tabletypesize{\small}
\tablecaption{Results on Continuum.\label{tab:results_dust}}
\tablehead{
\colhead{GRB} & \colhead{$\lambda_{\rm obs}$$^a$} & \colhead{$S_{\rm cont}$$^b$} & 
\colhead{$M_{\rm dust}$$^c$} & \colhead{$L_{\rm FIR}$$^c$} & \colhead{SFR$^c$} & \colhead{$M_{\rm gas}/M_{\rm dust}$$^d$} \\
\colhead{} & \colhead{(mm)} & \colhead{($\mu$Jy)} & 
\colhead{($M_{\odot}$)} & \colhead{($L_{\odot}$)} & \colhead{($M_{\odot}$~yr$^{-1}$)} & \colhead{} 
}
\startdata
980703 &1.3&$<$81    &$<$$1.8 \times 10^7$    &$<$$2.4 \times 10^{11}$    & $<$25   & --           \\
021211 &1.4&$<$63    &$<$$1.5 \times 10^7$    &$<$$2.0 \times 10^{11}$    & $<$21   & --           \\
031203 &1.0&$<$2410  &$<$$1.4 \times 10^7$    &$<$$1.8 \times 10^{11}$    & $<$19   & --           \\
050826 &1.2&$<$84    &$<$$4.1 \times 10^6$    &$<$$5.5 \times 10^{10}$    & $<$5.7  & $>$720       \\
051006 &1.3&$78\pm24$&$(2.1\pm0.6)\times10^7$ &$(2.8\pm0.9)\times10^{11}$ &$29\pm9$ & $1730\pm560$ \\
051117B&1.3&$69\pm20$&$(9.2\pm2.7)\times10^6$ &$(1.2\pm0.4)\times10^{11}$ &$13\pm4$ & $640\pm260$  \\
060814 &2.0&$<$63    &$<$$5.3 \times 10^7$    &$<$$7.1 \times 10^{11}$    & $<$74   & $>$1070      \\
070306 &2.1&$<$36    &$<$$4.8 \times 10^7$    &$<$$6.4 \times 10^{11}$    & $<$68   & $>$600       \\
071021 &2.1&$<$42    &$<$$5.4 \times 10^7$    &$<$$7.2 \times 10^{11}$    & $<$76   & --           \\
081109A&1.3&$<$87    &$<$$2.0 \times 10^7$    &$<$$2.7 \times 10^{11}$    & $<$28   & $>$1000      \\
100621A&1.3&$<$66    &$<$$1.1 \times 10^7$    &$<$$1.5 \times 10^{11}$    & $<$16   & --           \\
110918A&1.3&$97\pm20$&$(2.2\pm0.5)\times10^7$ &$(3.0\pm0.6)\times10^{11}$ &$32\pm7$ & $840\pm210$  \\
130925A&1.1&$<$96    &$<$$6.5 \times 10^6$    &$<$$8.8 \times 10^{10}$    & $<$9.2  & --           \\
140301A&2.2&$<$39    &$<$$4.7 \times 10^7$    &$<$$6.3 \times 10^{11}$    & $<$66   & $>$920       \\
\enddata
\tablecomments{
Limits are 3$\sigma$. 
$^a$ Representative observed wavelength for continuum. 
$^b$ Continuum flux. 
$^c$ Dust mass, FIR luminosity, and SFR derived from the continuum flux by assuming $T_{\rm dust} = 40$~K. 
$^d$ Molecular gas mass to dust mass ratio.
}
\end{deluxetable*}

\subsection{Individual Hosts} \label{sec:individual}

\subsubsection{GRB~980703}

The GRB~980703 host is an actively star-forming galaxy with a UV-based SFR derived from SED fitting of $\sim$10--40 $M_{\odot}$~yr$^{-1}$ \citep{cast10, sven10, perl13}. 
The host was detected in radio observations at 1.43, 4.86, and 8.46 GHz, and the radio-based SFR is $110 \pm 15$~$M_{\odot}$~yr$^{-1}$ \citep{berg01, berg03}. 
Recent radio follow-up observations ($\gtrsim$14 years after the GRB) detected the host with fainter flux densities than those reporeted by \cite{berg01, berg03}, providing updated radio-based SFRs of $77 \pm 22$ and $93 \pm 21$~$M_{\odot}$~yr$^{-1}$ at 6 and 1.45 GHz, respectively \citep{perl17a}. 
They suggested an afterglow contribution to the radio fluxes in the previous studies based on long-term radio observations and light curve modeling.

The CO(4--3) line and continuum were not detected, although a marginal feature of CO emission can be seen. 
The upper limit on $L_{\rm IR}$ based on the continuum emission is $<$$6 \times 10^{11}$ $L_{\odot}$, suggesting that the host does not have a ULIRG nature. 
The upper limit on SFR is $<$25~$M_{\odot}$~yr$^{-1}$, which is lower than the results of early radio observations by \cite{berg03}. 
Although the UV-to-radio SED fitting by \cite{mich08} and \cite{hunt14} showed a large SFR of 90--130 $M_{\odot}$~yr$^{-1}$, this could be overestimated because they adopted the radio flux densities of \cite{berg03}, and the estimated 1.3~mm flux density was larger than our ALMA upper limit on continuum emission.

\subsubsection{GRB~021211}

The UV-based SFR from the SED fitting of the GRB~021211 host showed an SFR of 0.4--8~$M_{\odot}$~yr$^{-1}$ \citep{sava09, cast10, sven10, perl13}. 
\cite{mich12} detected 1.43 GHz radio emission from the host with VLA. 
The radio-based SFR is $\sim$500 $M_{\odot}$~yr$^{-1}$, placing it in the category of ULIRGs. 
However, the Australia Telescope Compact Array (ATCA) 2 GHz observations by \cite{hats12} did not detect emission, and the authors placed an upper limit on SFR $< 41$~$M_{\odot}$~yr$^{-1}$. 
Re-observations with VLA by \cite{perl17a} also failed to detect emission at 1.4 and 2.1 GHz. 
They re-analyzed the original data taken by \cite{mich12} and found that no source is detected at the afterglow position, concluding that the previously reported detection was likely a processing artifact.

Our ALMA observations did not detect the CO(4--3) line and continuum emission. 
The derived upper limits on IR luminosity and SFR ($L_{\rm IR} < 2.0 \times 10^{11}$~$L_{\odot}$ and SFR $< 21$~$M_{\odot}$~yr$^{-1}$) are consistent with the radio non-detection \citep{hats12, perl17a}.

\subsubsection{GRB~031203}

GRB~031203 at $z = 0.1055$ is one of the closest long-duration GRBs, allowing us to study the environment of a GRB in detail. 
The host is a dwarf galaxy with a stellar mass of $3 \times 10^8$~$M_{\odot}$ \citep{guse11, perl13}. 
The UV-based SFR from SED fitting is 0.3--14~$M_{\odot}$~yr$^{-1}$ \citep{cast10, sven10, perl13}, and the H$\alpha$-based SFR is 3--16~$M_{\odot}$~yr$^{-1}$ \citep{proc04, marg07, sava09, leve10a, guse11}. 
{\sl Herschel}/PACS observations by \cite{syme14} tentatively detected the host at 70~$\mu$m ($\sim$$2.2\sigma$) and derived $\log M_{\rm dust} = 4.27$~$M_{\odot}$, $\log L_{\rm IR} = 10.47$~$L_{\odot}$, and SFR $= 3$~$M_{\odot}$~yr$^{-1}$ from the IR SED. 
Radio observations with ATCA detected the host at 1.39, 2.37, and 5.5 GHz \citep{stan10, mich10} and a radio-based SFR of 2--3~$M_{\odot}$~yr$^{-1}$ was derived, which is consistent with previous studies.

Our ALMA observations did not detect the CO(3--2) line and continuum emission. 
The upper limit on molecular gas mass is $3.5 \times 10^9$~$M_{\odot}$, which is consistent with the upper limit of $9.8 \times 10^9$~$M_{\odot}$ reported by \cite{mich18} derived from the CO(2--1) line observations. 
\cite{wier18} derived an H$_2$ gas mass (including He and heavy elements) of $2.3 \times 10^9$~$M_{\odot}$ based on the tentative detection of the rotation-vibrational H$_2$ 0--0 S(7) line. 
They assumed that the column density of H$_2$ molecules is distributed as a power-law function with respect to temperature, with an upper temperature of 2000~K, a lower temperature of 50~K, and a power-law index of $n = 4.5$. 
The derived H$_2$ mass is consistent with our upper limit.

The dust continuum emission was not detected, placing upper limits of 
$M_{\rm dust} < 1.4 \times 10^7$~$M_{\odot}$, $L_{\rm IR} < 1.8 \times 10^{11}$~$L_{\odot}$, and SFR $< 19$~$M_{\odot}$~yr$^{-1}$, which are consistent with the results of previous studies.

\subsubsection{GRB~050826}

The GRB~050826 host has an SFR of 1--2~$M_{\odot}$~yr$^{-1}$ derived from UV-SED and H$\alpha$ emission \citep{sven10, leve10a}.

The CO(3--2) line was significantly detected and the derived molecular gas mass is $(2.9 \pm 0.5) \times 10^9$~$M_{\odot}$. 
The velocity-integrated intensity map shows a possible companion (S/N $\sim 3$) $\sim$$1''$ west of the host with a velocity offset of $\sim$100 km~s$^{-1}$. 
The two components appeared to be connected with weak CO emission, which might indicate an interaction.

The dust continuum was not detected, placing upper limits of 
$M_{\rm dust} < 4.1 \times 10^6$~$M_{\odot}$, $L_{\rm IR} < 5.5 \times 10^{10}$~$L_{\odot}$, and SFR $<5.7$~$M_{\odot}$~yr$^{-1}$. 
The SFR upper limit is consistent with previous studies.

\subsubsection{GRB~051006}

Because of the limited early optical observations, it is not known whether GRB~051006 is a dark GRB or not \citep{perl15}. 
The host was detected in VLA 3 GHz observations and the radio-based SFR is $51^{+22}_{-18}$~$M_{\odot}$~yr$^{-1}$ \citep{perl15}. 
The SED fitting also shows a high SFR of $98^{+2}_{-1}$~$M_{\odot}$~yr$^{-1}$ \citep{perl15}.

The CO(4--3) line was significantly detected, and the derived molecular gas mass is ($3.6 \pm 0.4) \times 10^{10}$~$M_{\odot}$. 
The dust continuum was detected with S/N $= 3.3$ and the estimated SFR is $29 \pm 9$ $M_{\odot}$~yr$^{-1}$. 
The FIR-based SFR is lower than the results of our SED analysis or \cite{perl15}. 
This might be because the dust temperature is higher than that we assumed (40 K).

\subsubsection{GRB~051117B}

The GRB~051117B host is known to be one of the highest-metallicity GRB hosts, with a super-solar metallicity with $12 + \log{\rm (O/H)} = 9.00 \pm 0.16$ \citep{kruh15}, which was derived with ${\rm N2}$ and ${\rm O3N2}$ diagnostics based on the calibrators of \cite{naga06} and \cite{maio08}. 
The UV-based SFR derived from SED fitting is 3.4 $M_{\odot}$~yr$^{-1}$ \citep{kruh17} and the H$\alpha$-based SFR is $4.7^{+4.9}_{-2.2}$~$M_{\odot}$~yr$^{-1}$ \citep{kruh15}. 
The radio observations at 3, 5.5, and 9 GHz did not detect emission \citep{mich12, perl15}, giving an upper limit of SFR $< 27$~$M_{\odot}$~yr$^{-1}$ \citep{mich12}.

The CO(3--2) line was detected and the molecular gas mass is ($5.9 \pm 1.7) \times 10^9$~$M_{\odot}$. 
The dust continuum was detected (S/N $\sim 3.4$). 
The estimated SFR is $13 \pm 4$ $M_{\odot}$~yr$^{-1}$, which is consistent with the H$\alpha$-derived SFR and the upper limits obtained in the radio observations.

\subsubsection{GRB~060814}

GRB~060814 is a dark GRB \citep{kruh12}. 
The host complex consists of components A and B (spectroscopic redshifts of $z = 1.923$ and 1.920, respectively) with a separation of $1\farcs3$ and the GRB occurs at component A \citep{jako12, kruh12}. 
The host shows active star formation, with SFR of $\sim$50~$M_{\odot}$~yr$^{-1}$ \citep{kruh15, palm19} derived from H$\alpha$ line, 210--240~$M_{\odot}$~yr$^{-1}$ derived from SED modeling \citep{perl13, perl15}, and 260~$M_{\odot}$~yr$^{-1}$ derived from 3 GHz radio observations \citep{perl15}.

We detected the CO(3--2) line at $z = 1.923$, which is consistent with the redshift of the component A. 
The molecular gas mass is $(5.7 \pm 1.1) \times 10^{10}$~$M_{\odot}$, which is one of the most massive GRB hosts \citep{hats19}. 
In spite of the large SFR, the dust continuum was not detected in our observations, giving an upper limit of SFR $<74$~$M_{\odot}$~yr$^{-1}$.

\subsubsection{GRB~070306}

The afterglow of GRB~070306 is highly extinguished, with a visual extinction of $A_V = 5.5$~mag, and the GRB is classified as a dark burst \citep{jaun08}. 
SED modeling of the host by \cite{perl13} showed a relatively blue SED with a low mean extinction ($A_V = 0.13$), suggesting a complex dust geometry in the host. 
The UV-based SFR derived from SED fitting is 10--20~$M_{\odot}$~yr$^{-1}$ \citep{kruh11, perl13, perl15}, whereas the SFR derived from the IR SED with {\sl Herschel}/PACS photometry is 100--140~$M_{\odot}$~yr$^{-1}$ \citep{hunt14, scha14}. 
The SFR derived from 3 GHz radio observations is $143^{+61}_{-35}$~$M_{\odot}$~yr$^{-1}$ \citep{perl15}, which is consistent with the IR SFR and higher than the UV-based SFR, suggesting dust obscured star formation in the host.

We detected the CO(3--2) line and derived a molecular gas mass of $(2.9 \pm 0.5) \times 10^{10}$~$M_{\odot}$. 
The dust continuum was not detected, providing an upper limit of SFR $<68$~$M_{\odot}$~yr$^{-1}$.

\subsubsection{GRB~071021}

GRB~071021 is a dark GRB \citep{kruh12}. 
The host at $z = 2.452$ exhibits the highest redshift in our sample. 
The host is an actively star-forming galaxy with SFR $= 32^{+20}_{-12}$~$M_{\odot}$~yr$^{-1}$ derived from H$\alpha$ line \citep{kruh15} and $190.3^{+25.6}_{-20.3}$~$M_{\odot}$~yr$^{-1}$ derived from SED modeling \citep{perl13}.

In spite of the large SFR, neither the CO(4--3) line nor the dust continuum emission were detected. 
Note that the bright continuum source $2''$ southeast of the GRB is an unrelated object. 
The derived upper limits are $M_{\rm gas} < 4.6 \times 10^{10}$~$M_{\odot}$ and SFR $<76$~$M_{\odot}$~yr$^{-1}$.

\subsubsection{GRB~081109A}

GRB~081109A is a dark GRB, and the afterglow spectrum shows a high visual extinction \citep[$A_V = 3.4$~mag;][]{kruh11}. 
The host galaxy shows active star-forming properties, with SFR $\sim 10$--50~$M_{\odot}$~yr$^{-1}$ based on the SED modeling and emission lines of H$\alpha$ and [OII] \citep{kruh11, perl13, kruh15}.

We detected the CO(4--3) line and obtained a molecular gas mass of $(2.0 \pm 0.3) \times 10^{10}$~$M_{\odot}$. 
The CO velocity field shows a rotation feature in the southwest-northeast direction. 
The dust continuum was not detected, and the derived upper limit is SFR $<28$~$M_{\odot}$~yr$^{-1}$.

\subsubsection{GRB~100621A}

GRB~100621A is another dark GRB, and the afterglow shows high visual extinction \citep[$A_V = 3.8$~mag;][]{kruh11}. 
However, the host shows blue SED \citep{kruh11}. 
The red afterglow and blue host suggest a complex dust geometry in the host, as in the case of GRB~070306 \citep{kruh11, perl13}. 
The SFR based on SED modeling or the H$\alpha$ line is $\sim$10~$M_{\odot}$~yr$^{-1}$ \citep{kruh11, kruh15, jape16}. 
Although \cite{stan14} reported radio detection at 5.5 and 9.0 GHz $\sim$0.8 year after the explosion and a derived SFR of 60~$M_{\odot}$~yr$^{-1}$, the follow-up observations at 2.1 GHz by \cite{grei16} $\sim$2.7 years after the explosion did not confirm the detection. 
The obtained upper limit is SFR $< 30$~$M_{\odot}$~yr$^{-1}$. 
In combination with earlier radio flux measurements of the afterglow, \cite{grei16} suggested that the radio emission reported in \cite{stan14} was due to the afterglow.

The host was not detected in either the CO(3--2) line or the dust continuum in our observations. 
The upper limits are $M_{\rm gas} < 1.9 \times 10^9$~$M_{\odot}$ and SFR $<16$~$M_{\odot}$~yr$^{-1}$.

\subsubsection{GRB~110918A}

The afterglow of GRB~110918A shows low extinction ($A_V = 0.16$~mag), whereas the host is massive and metal-rich \citep{elli13}. 
The host has a solar metallicity with $12 + \log{\rm (O/H)} = 8.93 \pm 0.11$ \citep{kruh15}. 
The SFR is 20--40 $M_{\odot}$~yr$^{-1}$ based on the H$\alpha$ line \citep{elli13, kruh15} and 66~$M_{\odot}$~yr$^{-1}$ based on SED fitting \citep{elli13}. 
Radio 2.1 GHz observations by \cite{grei16} put an upper limit of SFR $< 84$~$M_{\odot}$~yr$^{-1}$.

The CO(3--2) line and dust continuum were detected at the host center, but they were not detected at the GRB position 12 kpc away from the host center \citep{elli13}. 
The CO velocity field shows a gradient consistent with rotation. 
The derived molecular gas mass is $M_{\rm gas} = (1.9 \pm 0.3) \times 10^9$~$M_{\odot}$. 
The FIR-based SFR is $32 \pm 7$ $M_{\odot}$~yr$^{-1}$, which consistent with the previous results.

\subsubsection{GRB~130925A}

GRB~130925A is an ultra-long GRB with a duration of prompt gamma-ray emission of $\sim 20$ ks \citep{bell14, evan14, piro14}. 
The afterglow spectrum shows a high visual extinction \citep[$A_V = 5.0 \pm 0.7$ mag;][]{grei14} that is one of the largest extinction among GRB afterglows. 
The host also shows a high extinction of $A_V = 2.4 \pm 0.9$ mag based on SED fitting \citep{scha15}. 
Optical spectroscopy of the host shows two star-forming regions corresponding to the galaxy nucleus and an {\sc Hii} region located $\sim$0\farcs2 (6 kpc in projection) southwest of the nucleus \citep{scha15}. 
The GRB site is close to the nucleus, which has a super-solar metallicity \citep{scha15, kruh15}. 
The SFR in the host is 2--3~$M_{\odot}$~yr$^{-1}$ based on the H$\alpha$ line and SED fitting \citep{scha15, kruh15}.

Our observations did not detect either the CO(3--2) line or the dust continuum, placing upper limits of $M_{\rm gas} < 7.6 \times 10^8$ $M_{\odot}$ and SFR $<9.2$~$M_{\odot}$~yr$^{-1}$.

\subsubsection{GRB~140301A}

The host has a high SFR of $106^{+36}_{-25}$~$M_{\odot}$~yr$^{-1}$ based on the H$\alpha$ line \citep{kruh15}. 
The host is one of the highest metallicity GRB hosts, with $12 + \log{\rm (O/H)} = 8.89 \pm 0.09$ \citep{kruh15}.

The CO(3--2) line was significantly detected, whereas the dust continuum was only tentatively detected (S/N $= 2.9$). 
The CO velocity field shows a gradient consistent with rotation. 
The derived molecular gas mass is $M_{\rm gas} = (4.3 \pm 0.5) \times 10^{10}$~$M_{\odot}$. 
The upper limit on SFR is $<$$66$ $M_{\odot}$~yr$^{-1}$, which is lower than the results of the H$\alpha$ observations and our SED analysis. 
A possible cause of the difference may be the assumed dust temperature, where a larger SFR would be derived for a higher dust temperature.

\section{Discussion}\label{sec:discussion}

\subsection{Line Width}\label{sec:fwhm}

\begin{figure}
\begin{center}
\includegraphics[width=\linewidth]{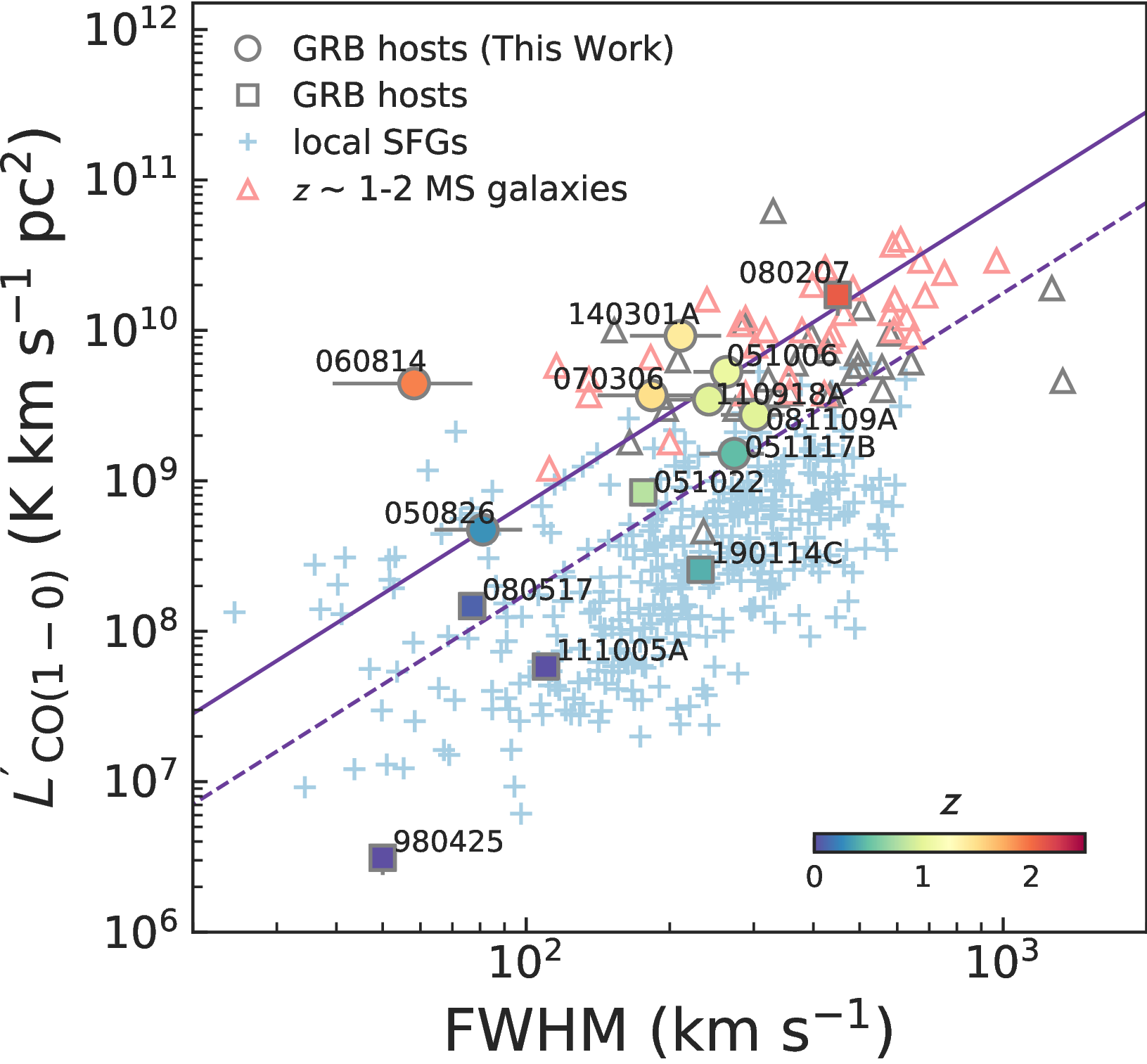}
\end{center}
\caption{
CO(1-0) line luminosities as a function of the line widths. 
Circles and squares represent GRB hosts of our targets and in the literature, respectively. 
For comparison, we plot $z \sim 1$--2 MS galaxies \citep{dadd10, magn12, magd12, tacc13} and $z = 0.01$--0.05 star-forming galaxies \citep{both14, sain17}.
The comparison sample with SFRs below the target selection criteria are shown in gray. 
The lines represent a formula for a disk model with $R = 4$~kpc and $\alpha_{\rm CO} = 5$ (solid) and with $R = 2$~kpc and $\alpha_{\rm CO} = 10$ (dashed), respectively. 
The GRB hosts are color coded by redshift. 
}
\label{fig:fwhm-lco}
\end{figure}

The CO line luminosity and the line FWHM are indicators of the molecular gas mass and the dynamical mass, respectively. 
In Figure~\ref{fig:fwhm-lco} we compare $L'_{\rm CO}$ and FWHM for CO-detected GRB hosts in our sample and in the literature. 
We also show the sample of $z \sim 1$--2 MS galaxies \citep{dadd10, magn12, magd12, tacc13} and $z = 0.01$--0.05 star-forming galaxies \citep{both14, sain17}. 
The comparison sample is limited to have the same selection criteria on SFR as the tergets (Section~\ref{sec:targets}). 
The GRB hosts at low redshifts share a similar region to local star-forming galaxies. 
The GRB hosts at $z \sim 1$--2 overlap the MS galaxies at similar redshifts except for the GRB~060814 host. 
It might be possible that the line width would be broader if we take the emission from $\sim$$-180$ to 180 km~s$^{-1}$ as seen in Figure~\ref{fig:results}, although the emission is not significant. 
Another possibility is that a small inclination angle make the line width narrower.

We presents a relation between $L'_{\rm CO}$ and FWHM following \cite{both13}: 
\begin{equation}
L'_{\rm CO} = C \left( \frac{R}{\alpha_{\rm CO} G} \right) \left( \frac{\Delta v_{\rm FWHM}}{2.35} \right), 
\end{equation}
where $C$ is a constant that depends on the mass distribution and kinematics of the galaxy \citep{erb06}, $R$ is the radius of the CO-emitting region in kpc, $G$ is the gravitational constant, and $\Delta v_{\rm FWHM}$ is the line width in km~s$^{-1}$. 
We take $C = 2.1$ for a disk galaxy following previous studies on $z > 1$ star-forming galaxies and submillimeter galaxies (SMGs) \citep{both13, arav16}. 
We note that there are many possible uncertainties both from the assumed parameters and observational measurements. 
Nevertheless, as shown in Figure~\ref{fig:fwhm-lco}, the GRB hosts appear to follow the model relation, which is also applied to $z \sim 1$--2 MS galaxies, suggesting a similarity between the two populations in terms of geometry and kinematics.

\subsection{Molecular Gas Mass--SFR}\label{sec:mgas-sfr}

\begin{figure}
\begin{center}
\includegraphics[width=\linewidth]{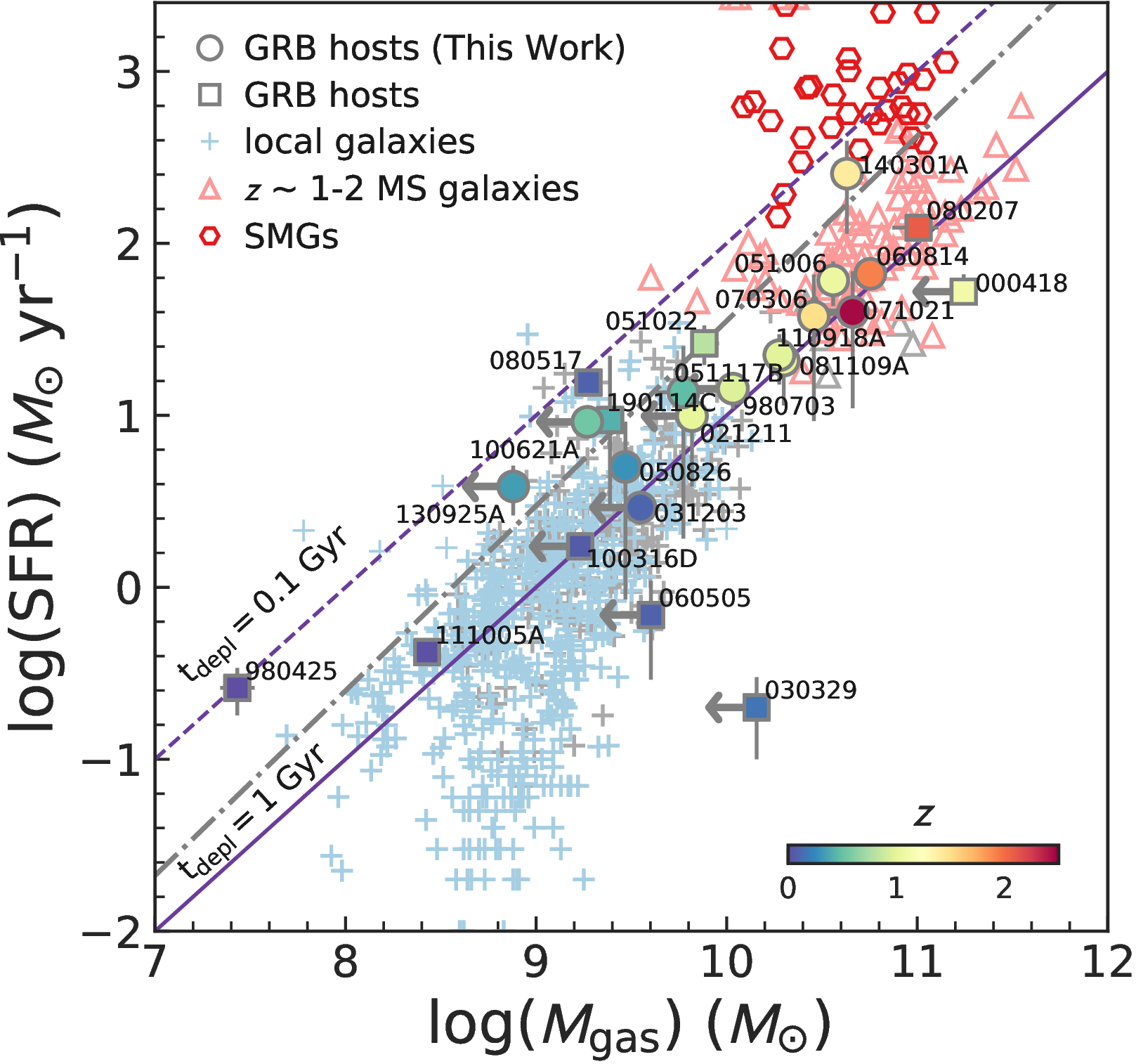}
\end{center}
\caption{
Comparison of molecular gas mass and SFR. 
We plot the GRB host galaxies in our sample and in the literature. 
The GRB hosts are color coded by redshift. 
The SFRs of our targets are derived by the SED fitting. 
For comparison, we plot 
local galaxies \citep{sain11, sain17, both14}, 
$z \sim 1$--2 MS galaxies \citep{tacc13, seko16a}, 
and SMGs \citep{both13}. 
The comparison sample with SFRs below the target selection criteria are shown in gray.
The solid and dashed lines represent gas depletion times of 0.1 and 1 Gyr, respectively. 
The typical 5$\sigma$ detection limit is shown as a dot-dashed line by assuming a Galactic CO-to-H$_2$ conversion factor of 4.4 $M_{\odot}$~(K~km~s$^{-1}$~pc$^2$)$^{-1}$. 
}
\label{fig:mgas-sfr}
\end{figure}

There is a correlation between gas surface density and SFR surface density known as the Kennicutt--Schmidt relation \citep{kenn98, schm59}. 
It is also known that their integrations over the galaxy show a correlation ($M_{\rm gas}$--SFR). 
Figure~\ref{fig:mgas-sfr} compares the molecular gas masses and SFRs of the targets. 
The SFRs are derived by the SED fitting (Section~\ref{sec:sed}). 
We also plot the GRB hosts with CO observations in the literature by using the same CO line ratios and metallicity-dependent conversion factors adopted for our sample. 
For comparison, local galaxies \citep{sain11, sain17, both14}, $z \sim 1$--2 MS galaxies \citep{tacc13, seko16a}, and SMGs \citep{both13} are also plotted. 
Since the location in the $M_{\rm gas}$--SFR plot might be biased for a galaxy above the MS line, we limit the comparison sample with the same selection criteria on SFR as our targets (Section~\ref{sec:targets}).

The majority of the GRB hosts at $z \gtrsim 1$ are located in the region similar to those of $z \sim 1$--2 MS galaxies with a molecular gas depletion timescale ($t_{\rm depl} = M_{\rm gas}/{\rm SFR}$) of $\sim$1 Gyr, while some hosts at lower redshifts have a shorter gas depletion timescale. 
The upper limit on molecular gas mass on the GRB~100621 and 130925A hosts places them in the sequence of starburst galaxies, where the hosts of GRBs 980425 and 080517 are also located \citep{stan15, mich16}. 
It is possible that these hosts are in the burst phase of star-formation with an enhanced star-formation efficiency (SFR/$M_{\rm gas}$). 
Although \cite{mich18} reported that the hosts of GRBs 060814 and 100316D are deficient in molecular gas, we do not find the deficiency in Figure~\ref{fig:mgas-sfr}. 
This is because they adopted a fixed value of conversion factor $\alpha_{\rm CO} = 5$ $M_{\odot}$~(K~km~s$^{-1}$~pc$^2$)$^{-1}$, whereas we adopt a metallicity-dependent $\alpha_{\rm CO}(Z)$, which is larger than 5 $M_{\odot}$~(K~km~s$^{-1}$~pc$^2$)$^{-1}$ (Table \ref{tab:other_hosts}).

\begin{figure*}
\begin{center}
\includegraphics[width=.46\linewidth]{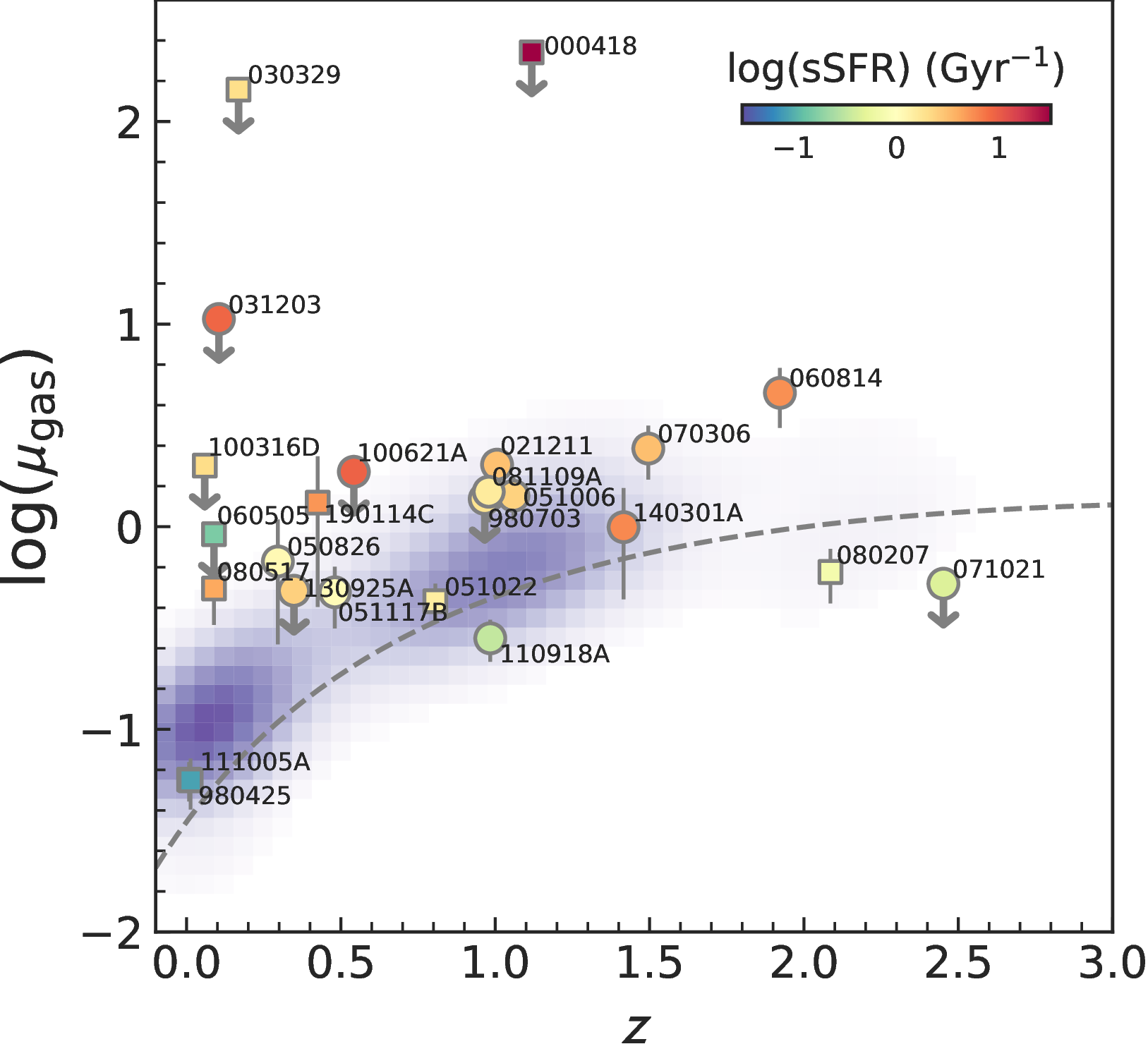}
\hspace{3mm}
\includegraphics[width=.48\linewidth]{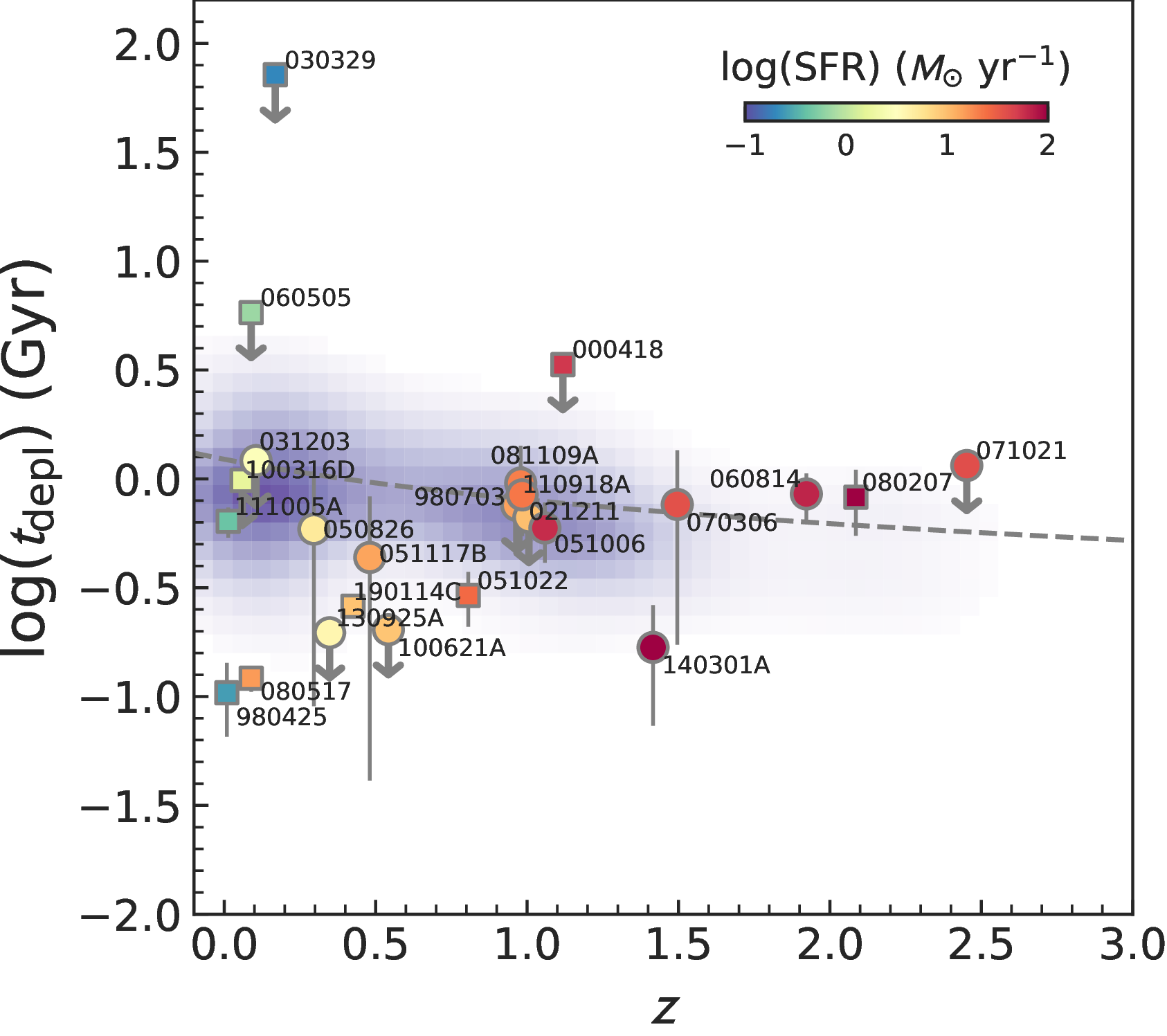}
\end{center}
\caption{
Molecular gas fraction ($\mu_{\rm gas} = M_{\rm gas}/M_*$) (left) and molecular gas depletion timescale ($t_{\rm depl} = M_{\rm gas}/{\rm SFR}$) (right) as a function of redshift. 
Circles and squares represent GRB hosts of our targets and in the literature, respectively. 
The background distribution shows the distribution of a sample of star-forming galaxies compiled by \cite{tacc18}. 
The dashed line shows the best-fit line for the MS galaxies derived in \cite{tacc18}. 
The GRB hosts are color coded by sSFR (left panel) and SFR (right panel). 
}
\label{fig:z-mugas}
\end{figure*}

\subsection{Gas Fraction, Depletion Timescale, and Scaling Relation}\label{sec:scaling}

Previous studies on GRB hosts suggest a deficiency of molecular gas in the host galaxies compared to their star formation rate (SFR) or stellar mass \citep{hats14, stan15, mich16}. 
\cite{stan15} claimed that the CO emission of GRB hosts to be weak compared with their SFRs, suggesting that a short gas consumption timescale is characteristic of GRB hosts. 
However, recent CO observations by \cite{arab18} and \cite{mich18} found no molecular gas deficiency, and the reported deficiency can be due to the small statistics and/or the adopted CO-to-H$_2$ conversion factor. 
CO(1--0) observations of the GRB~080207 host by \cite{hats19} support the claim that GRBs hosts can be representative star-forming galaxies.

We examine this issue with the largest sample of GRB hosts by comparing them with other star-forming galaxies. 
Figure~\ref{fig:z-mugas} shows the molecular gas fraction and molecular gas depletion timescale as a function of redshift. 
GRB hosts in the literature are also plotted by using the same CO line ratios and metallicity-dependent $\alpha_{\rm CO}$ adopted in this study. 
For comparison, the distribution of star-forming galaxies compiled by \cite{tacc18} is shown as a color distribution. 
They selected a representative and statistically significant sample of star-forming galaxies, covering a wide range in basic galaxy parameters and SFRs relative to that on the MS, and expanded the study of \cite{genz15} with a large sample of 1444 star-forming galaxies between $z =0$ and 4. 
The dashed line in Figure~\ref{fig:z-mugas} represents the best-fitting line for MS galaxies derived by \cite{tacc18}. 
The GRB hosts tend to have a higher $\mu_{\rm gas}$ and a shorter $t_{\rm depl}$ as compared with other star-forming galaxies at similar redshifts, especially at $z \lesssim 1$, although many of the data points for the GRB hosts are upper limits. 
This could be a common property of GRB hosts or an effect introduced by the selection of targets which are typically above the MS line.

\begin{figure*}
\centering
\includegraphics[width=.46\linewidth]{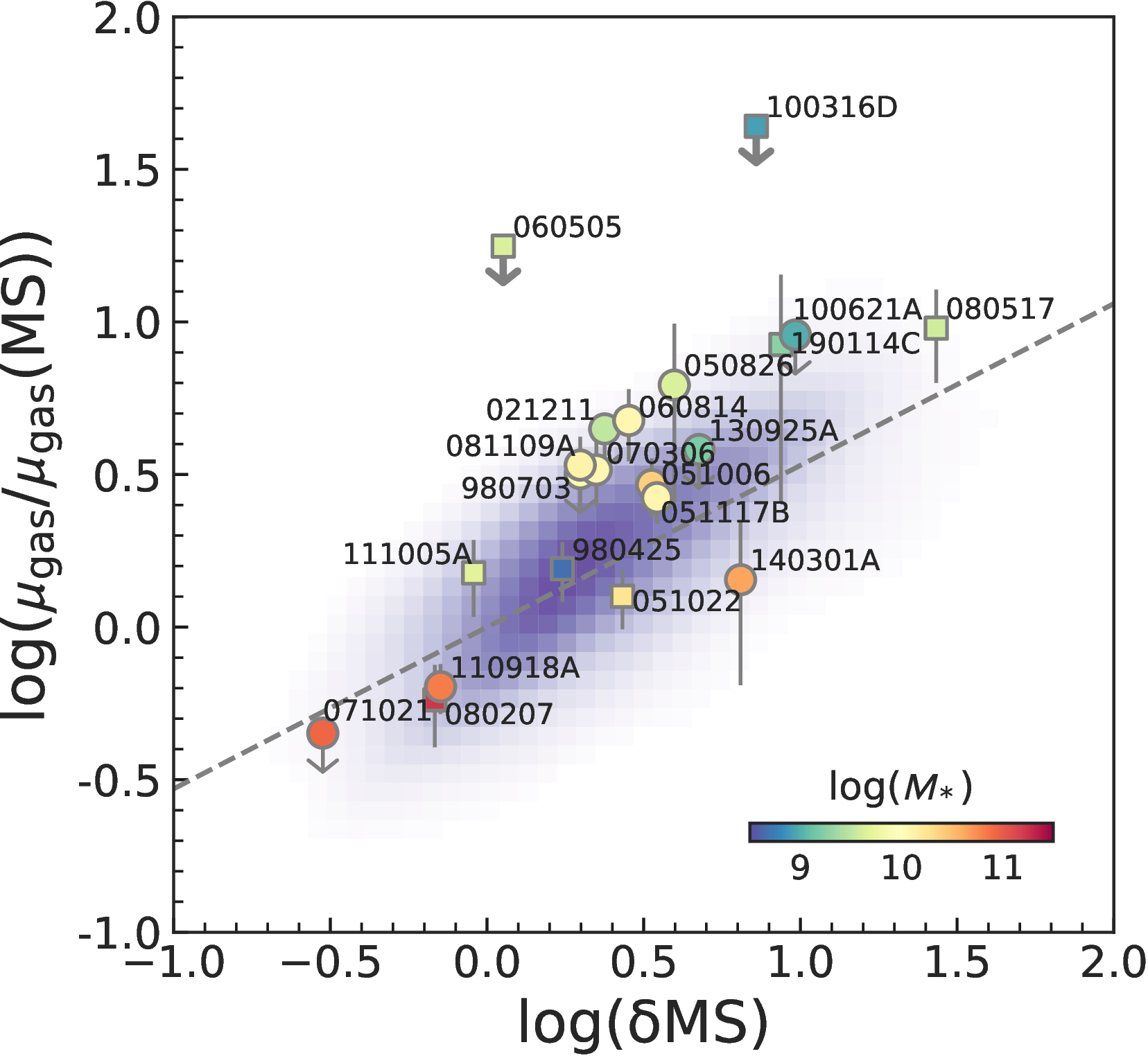}
\hspace{3mm}
\includegraphics[width=.47\linewidth]{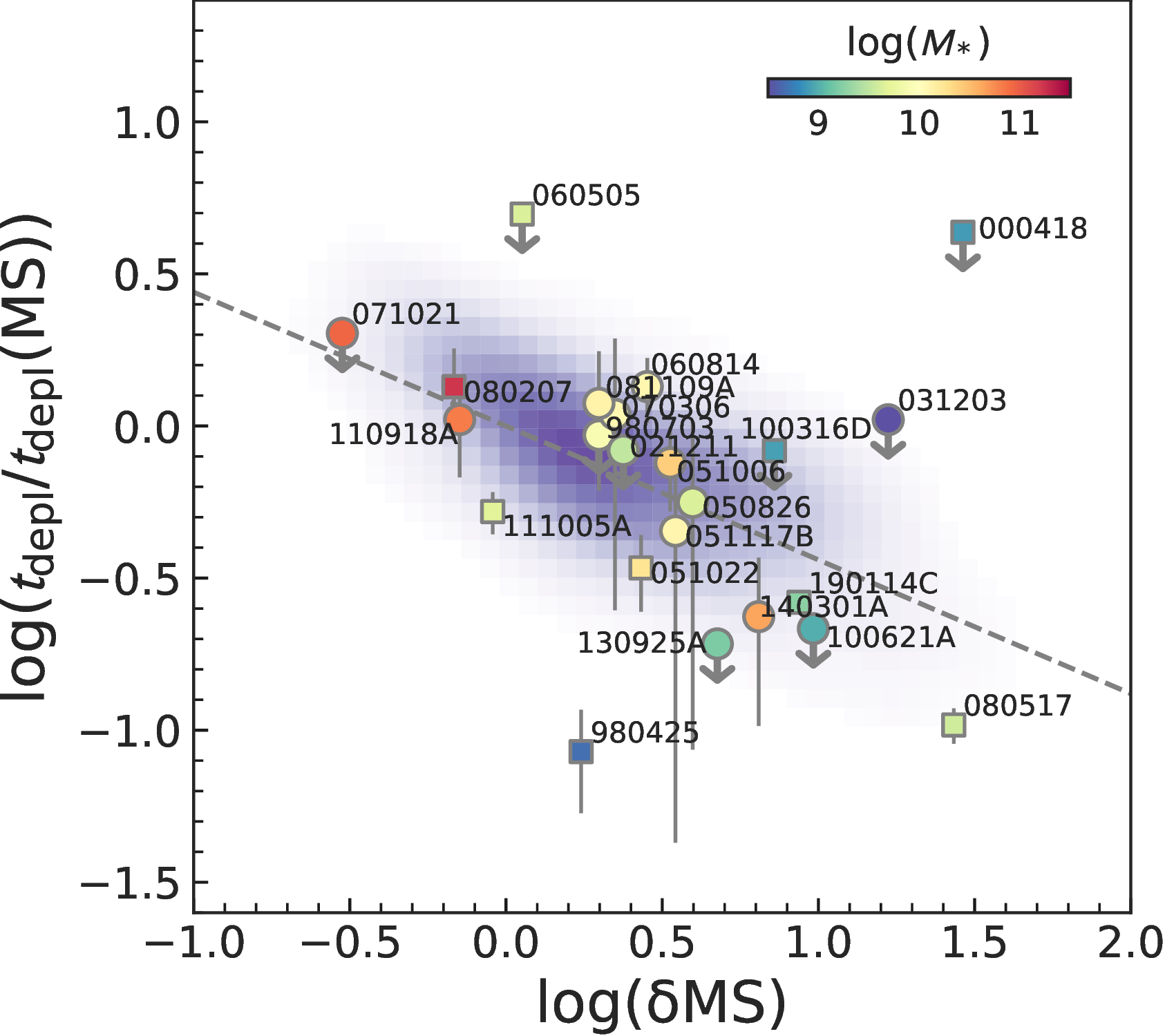}
\caption{
Dependence of molecular gas fraction ($\mu_{\rm gas}$) (left) and molecular gas depletion timescale ($t_{\rm depl}$) (right) as a function of the offset from the reference MS line ($\delta$MS). 
Circles and squares represent the GRB hosts of our targets and in the literature, respectively. 
The upper limit for the host of GRB~030329 (and GRB~000418 in the left panel) is not presented (above the plot range). 
The background distribution shows the distribution of a sample of star-forming galaxies compiled by \cite{tacc18}, and the dashed line shows the best-fit line for their sample. 
The GRB hosts are color coded by stellar mass. 
}
\label{fig:dms}
\vspace{10mm}
\end{figure*}

\begin{figure}
\centering
\includegraphics[width=\linewidth]{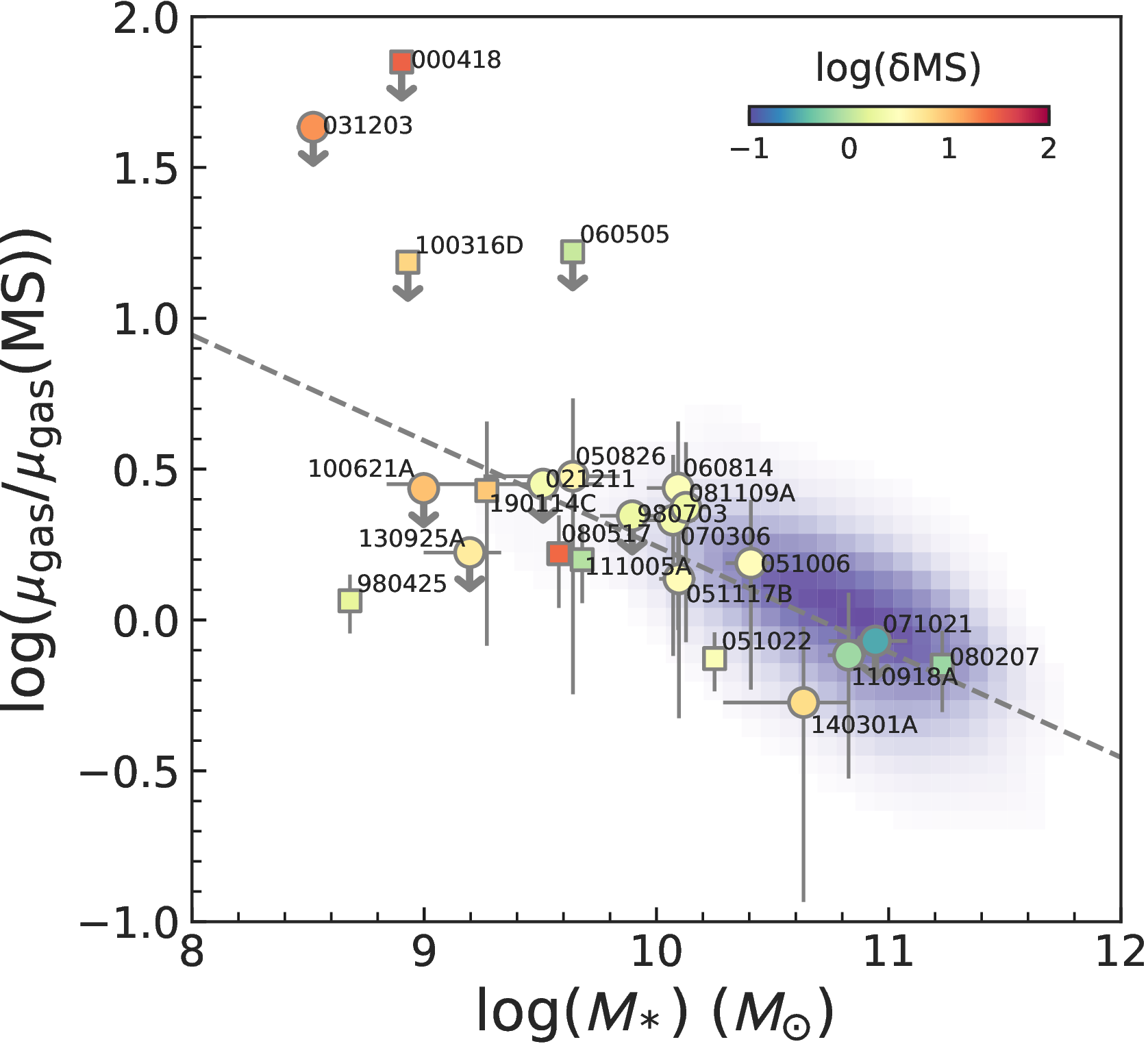}
\caption{
Dependence of molecular gas fraction ($\mu_{\rm gas}$) as a function of stellar mass. 
The symbols are the same as in Figure~\ref{fig:dms}. 
The GRB hosts are color coded by $\log(\delta$MS). 
}
\label{fig:mstar-mugasMS}
\end{figure}

\cite{genz15} and \cite{tacc18} argue the scaling relations for $\mu_{\rm gas}$ and $t_{\rm depl}$ can be written as functions depending on redshift, stellar mass, and offset from the MS line, $\delta$MS $=$ sSFR/sSFR(MS, $z, M_*$). 
\cite{tacc18} found that their large data sets follow the same scaling trends, where $\mu_{\rm gas}$ scales as $(1 + z)^{2.5} \times (\delta {\rm MS})^{0.52} \times (M_*)^{-0.36}$ and $t_{\rm depl}$ scales as $(1 + z)^{-0.6} \times (\delta {\rm MS})^{0.44}$ over a range of $\log \delta{\rm MS}$ from $-1$ to 2. 
The trend of the dependence on the distance from the MS line discussed in \cite{tacc18} is consistent with previous studies \citep[e.g.,][]{sain12, magd12, scov17}. 
Figure~\ref{fig:dms} shows $\mu_{\rm gas}$ and $t_{\rm depl}$ of the GRB hosts as a function of $\delta$MS. 
The GRB hosts follow the same scaling trends as other star-forming galaxies, where $\mu_{\rm gas}$ increases and $t_{\rm depl}$ decreases with increasing $\delta {\rm MS}$. 
The Pearson's correlation coefficients estimated for the CO-detected hosts are 
$r = 0.79$ and $r = -0.59$ for the distribution of $\mu_{\rm gas}$ and $t_{\rm depl}$, respectively, indicating a correlation. 
\cite{tacc18} found a dependence on stellar mass for $\mu_{\rm gas}$, and the trend of smaller $\mu_{\rm gas}$ for larger stellar mass also appears in the left panel of Figure~\ref{fig:dms} for the GRB hosts. 
This trend is presented in the plot of $\mu_{\rm gas}$ as a function of $\delta {\rm MS}$ in Figure~\ref{fig:mstar-mugasMS}. 
These suggest that GRB hosts have no molecular gas deficit when compared to other star-forming galaxies of similar SFR and stellar mass.

The GRB~980425 host is significantly offset from the trend in the $t_{\rm depl}$--$\delta$MS plot. 
The host is known to have a low CO-derived molecular gas mass for its SFR \citep{mich16}. 
There are also several possible reasons due to the uncertainties on the measured properties (stellar mass and SFR) in the literature, in addition to the assumed quantities ($\alpha_{\rm CO}$ and line ratios) in this study.

As discussed in Section~\ref{sec:fwhm}, the GRB hosts share dynamical properties with other star-forming galaxies. 
\cite{hats19} found that a $z = 2$ host of GRB~080207 shares common properties such as gas-to-dust ratio, location in the Kennicutt--Schmidt relation, and kinematics, in addition to $\mu_{\rm gas}$ and $t_{\rm depl}$ with normal star-forming galaxies at similar redshifts. 
The overall trend for the GRB hosts found in this study suggests that the star-forming environment producing GRBs is similar to that of other star-forming galaxies in terms of molecular gas. 
These suggest that the same star-formation mechanism is expected in GRB hosts as in other star-forming galaxies.

\section{Conclusions}\label{sec:conclusions}
We report the results of ALMA CO observations of 14 host galaxies of long-duration GRBs at $z = 0.1$--2.5. 
Eight hosts (GRBs 050826, 051006, 051117B, 060814, 070306, 081109, 110918A, and 140301A) at $z = 0.3$--2 were detected in CO [five with the CO(3--2) line and three with the CO(4--3) line], whereas only three hosts were detected in the dust continuum. 
Two hosts (GRBs 110918A and 140301A) show velocity gradient consistent with rotation with a line FWHM of 200--300 km s$^{-1}$. 
Molecular gas mass is estimated to be $M_{\rm gas} = (0.2$--$6) \times 10^{10}$~$M_{\odot}$ by assuming metallicity-dependent CO-to-H$_2$ conversion factors and CO line ratios appropriate for normal star-forming galaxies.

To estimate the SFR and stellar mass in a common way, we conducted SED analysis with available photometry from UV to radio including our ALMA photometry.

We combined the results of CO observations with those reported in the literature (11 GRB hosts) and discuss the CO and molecular gas properties with the largest sample of GRB hosts (25 in total, of which 14 are detected in CO). 
The key findings are as follows: 
\begin{itemize}
\item The CO-detected hosts share a similar region to local and $z \sim 1$--2 star-forming galaxies in the CO line FWHM--$L'_{\rm CO}$ plot, suggesting a similarity between the two populations in terms of geometry and kinematics.

\item 
The majority of the GRB hosts at $z \gtrsim 1$ are located at regions similar to those of $z \sim 1$--2 MS galaxies in the planes of $M_{\rm gas}$--$M_*$ and $M_{\rm gas}$--SFR.

\item 
The GRB hosts tend to have a higher $\mu_{\rm gas}$ and a shorter $t_{\rm depl}$ as compared with other star-forming galaxies at similar redshifts, especially at $z \lesssim 1$. 
This could be a common property of GRB hosts or an effect introduced by the selection of targets which are typically above the MS line.

\item 
In oder to eliminate the effect of selection bias toward higher-SFR hosts, we analyzed $\mu_{\rm gas}$ and $t_{\rm depl}$ as a function of the distance from the MS line ($\delta$MS). 
The GRB hosts were found to follow the same scaling relations as other star-forming galaxies derived by \cite{tacc18}, where $\mu_{\rm gas}$ increases and $t_{\rm depl}$ decreases with increasing $\delta {\rm MS}$.

\item 
When compared to other star-forming galaxies of similar SFR and stellar mass, no deficit of molecular gas is observed for the GRB hosts.
\end{itemize}

These findings suggest that the star-forming environment producing GRBs is similar to that of other star-forming galaxies in terms of molecular gas, especially molecular gas fraction and depletion timescale. 
This could indicate that the same star-formation mechanism is expected in GRB hosts as in other star-forming galaxies.

Because our targets for CO observations in this study were biased toward higher SFRs, it is important to construct an ``unbiased'' sample for studying the general properties of GRB hosts in future observations.

\acknowledgments

We would like to acknowledge the referee for helpful comments and suggestions. 
BH is supported by JSPS KAKENHI Grant Number 19K03925. 
This work is supported by a University Research Support Grant from the National Astronomical Observatory of Japan (NAOJ) and by the ALMA Japan Research Grant of NAOJ Chile Observatory (NAOJ-ALMA-230). 
This paper makes use of the following ALMA data: ADS/JAO.ALMA\#2015.1.00939.S and \#2016.1.00455.S. 
ALMA is a partnership of ESO (representing its member states), NSF (USA) and NINS (Japan), together with NRC (Canada), MOST and ASIAA (Taiwan), and KASI (Republic of Korea), in cooperation with the Republic of Chile. The Joint ALMA Observatory is operated by ESO, AUI/NRAO and NAOJ. 
This research has made use of the GHostS database (\url{www.grbhosts.org}), which is partly funded by {\sl Spitzer}/NASA grant RSA Agreement No. 1287913. 
 Based on observations made with the NASA/ESA Hubble Space Telescope, and obtained from the Hubble Legacy Archive, which is a collaboration between the Space Telescope Science Institute (STScI/NASA), the Space Telescope European Coordinating Facility (ST-ECF/ESA) and the Canadian Astronomy Data Centre (CADC/NRC/CSA).

This project used public archival data from the Dark Energy Survey (DES). Funding for the DES Projects has been provided by the U.S. Department of Energy, the U.S. National Science Foundation, the Ministry of Science and Education of Spain, the Science and Technology FacilitiesCouncil of the United Kingdom, the Higher Education Funding Council for England, the National Center for Supercomputing Applications at the University of Illinois at Urbana-Champaign, the Kavli Institute of Cosmological Physics at the University of Chicago, the Center for Cosmology and Astro-Particle Physics at the Ohio State University, the Mitchell Institute for Fundamental Physics and Astronomy at Texas A\&M University, Financiadora de Estudos e Projetos, Funda{\c c}{\~a}o Carlos Chagas Filho de Amparo {\`a} Pesquisa do Estado do Rio de Janeiro, Conselho Nacional de Desenvolvimento Cient{\'i}fico e Tecnol{\'o}gico and the Minist{\'e}rio da Ci{\^e}ncia, Tecnologia e Inova{\c c}{\~a}o, the Deutsche Forschungsgemeinschaft, and the Collaborating Institutions in the Dark Energy Survey.
The Collaborating Institutions are Argonne National Laboratory, the University of California at Santa Cruz, the University of Cambridge, Centro de Investigaciones Energ{\'e}ticas, Medioambientales y Tecnol{\'o}gicas-Madrid, the University of Chicago, University College London, the DES-Brazil Consortium, the University of Edinburgh, the Eidgen{\"o}ssische Technische Hochschule (ETH) Z{\"u}rich,  Fermi National Accelerator Laboratory, the University of Illinois at Urbana-Champaign, the Institut de Ci{\`e}ncies de l'Espai (IEEC/CSIC), the Institut de F{\'i}sica d'Altes Energies, Lawrence Berkeley National Laboratory, the Ludwig-Maximilians Universit{\"a}t M{\"u}nchen and the associated Excellence Cluster Universe, the University of Michigan, the National Optical Astronomy Observatory, the University of Nottingham, The Ohio State University, the OzDES Membership Consortium, the University of Pennsylvania, the University of Portsmouth, SLAC National Accelerator Laboratory, Stanford University, the University of Sussex, and Texas A\&M University.
Based in part on observations at Cerro Tololo Inter-American Observatory, National Optical Astronomy Observatory, which is operated by the Association of Universities for Research in Astronomy (AURA) under a cooperative agreement with the National Science Foundation.

\facility{ALMA}

\appendix
\section*{A. GRB Hosts with CO Observations in the Literature} \label{sec:other_hosts}




\begin{longrotatetable}
\begin{deluxetable*}{cccccccccccc}
\tablecaption{Properties of GRB hosts with CO observations in the literature} \label{tab:other_hosts}
\tablehead{
\colhead{GRB} & \colhead{$z$} & \colhead{SFR} & \colhead{$M_*$} & \colhead{Ref.} & 
\colhead{$12+\log({\rm O/H})$$^a$} & \colhead{Ref.} & 
\colhead{CO} & \colhead{$L'_{\rm CO}$$^b$} & \colhead{Ref.} & 
\colhead{$M_{\rm gas}$$^c$} & \colhead{$\alpha_{\rm CO}(Z)$$^d$} \\
\colhead{} & \colhead{} & \colhead{($M_{\odot}$~yr$^{-1}$)} & \colhead{($10^9 M_{\odot}$)} & \colhead{} & 
\colhead{} & \colhead{} & 
\colhead{} & \colhead{(K km s$^{-1}$~pc$^2$)} & 
\colhead{($M_{\odot}$)} & \colhead{} & \colhead{$M_{\odot}$~(K~km~s$^{-1}$~pc$^2$)$^{-1}$}
}
\startdata
980425 & 0.0085 & $0.26 \pm 0.08$        & $0.48 \pm 0.03$     & 1 & 8.31 &  8 & 2--1 & $2.3\times10^6$      & 16 & $2.7\times10^7$      &  8.7\\
000418 & 1.1183 & $52.4^{+13.7}_{-8.8}$  & $0.8^{+0.2}_{-0.1}$ & 2 & 8.16 &  9 & 2--1 & $<$$1.0\times10^{10}$& 17 & $<$$1.8\times10^{11}$& 12.7\\
030329 & 0.1685 & $0.2 \pm 0.1 $         & $0.1^{+0.0}_{-0.0}$ & 2 & 8.00 & 10 & 1--0 & $<$$6.9 \times 10^8$ & 18 & $<$$1.4\times10^{10}$& 20.8\\
051022 & 0.806  & $26.2^{+7.1}_{-6.6}$   & $17.8^{+3.8}_{-3.1}$& 2 & 8.33 & 11 & 4--3 & $4.2\times10^8$      & 19 & $8.7\times10^9$      &  7.3\\
060505 & 0.0889 & $0.69 \pm 0.40$        & 4.3                 & 3 & 8.41 & 12 & 2--1 & $<$$4.2 \times 10^8$ & 16 & $<$$2.0\times10^9$   &  7.2\\
060814 & 1.9229 & $238.2^{+49.6}_{-24.0}$& $9.8^{+0.9}_{-1.2}$ & 2 & 8.38 & 11 & 2--1 & $<$$8.3 \times 10^9$ & 16 & $<$$4.1\times10^{10}$&  7.5\\
080207 & 2.0858 & $123.4^{+25.2}_{-22.8}$& $170^{+8}_{-35}$    & 4 & 8.50 & 11 & 1--0 & $1.7 \times 10^{10}$ & 20 & $1.0\times10^{11}$   &  5.9\\
080517 & 0.089  & $15.5 \pm 0.5$         & $3.8 \pm 1.2$       & 5 & 8.33 & 13 & 1--0 & $1.5 \times 10^8$    & 21 & $1.2\times10^9$      & 12.8\\
100316D& 0.0591 & $1.73 \pm 0.08$        & 0.85                & 3 & 8.30 & 14 & 2--1 & $<$$1.4 \times 10^8$ & 16 & $<$$8.6\times10^8$   &  8.9\\
111005A& 0.01326& $0.42^{+0.06}_{-0.05}$ & $4.8^{+1.7}_{-0.9}$ & 6 & 8.63 & 15 & 2--1 & $4.4 \times 10^7$    & 16 & $2.7\times10^8$      &  4.6\\
190114C& 0.425  & $9.4^{+12.8}_{-6.4}$   & $1.9^{+1.7}_{-0.82}$& 7 & 8.27 &  7 & 3--2 & $1.54 \times 10^8$   &  7 & $2.4\times10^9$      &  9.5\\
\enddata
\tablecomments{
$^a$ Metallicity converted to the calibration of \cite{pett04} by using the metallicity conversion of \cite{kewl08}. 
$^b$ CO line luminosity. 
$^c$ Molecular gas mass with metallicity-dependent CO-to-H$_2$ conversion factor. 
$^d$ Metallicity-dependent CO-to-H$_2$ conversion factor. \\
(1) \citealt{mich14}; (2) \citealt{perl13}; (3) \citealt{mich15}; (4) \citealt{hash19}; (5) \citealt{stan15a}; (6) \citealt{mich18b}; 
(7) \citealt{deug19}; (8) \citealt{kruh17}; (9) \citealt{pira15}: (10) \citealt{leve10b}; (11) \citealt{kruh15}; (12) \citealt{thon14}; 
(13) \citealt{niin17}; (14) \citealt{leve11}; (15) \citealt{tang18}; 
(16) \citealt{mich18}; (17) \citealt{hats11}: (18) \citealt{endo07}: (19) \citealt{hats14}: (20) \citealt{hats19}; (21) \citealt{stan15}. 
}
\end{deluxetable*}
\end{longrotatetable}

\end{document}